# Consistency and prior falsification of training data in seismic deep learning: Application to offshore deltaic reservoir characterization

Anshuman Pradhan[*] and Tapan Mukerji
Department of Energy Resources Engineering
Stanford University
[*]pradhan1@stanford.edu

ABSTRACT

Deep learning applications of seismic reservoir characterization often require generation of synthetic data to augment available sparse labeled data. An approach for generating synthetic training data consists of specifying probability distributions modeling prior geologic uncertainty on reservoir properties and forward modeling the seismic data. A prior falsification approach is critical to establish the consistency of the synthetic training data distribution with real seismic data. With the help of a real case study of facies classification with convolutional neural networks (CNNs) from an offshore deltaic reservoir, we highlight several practical nuances associated with training deep learning models on synthetic seismic data. We highlight the issue of overfitting of CNNs to the synthetic training data distribution and propose regularization strategies to address it. We demonstrate the efficacy of our proposed strategies by training the CNN on synthetic data and making robust predictions with real 3D partial stack seismic data.

INTRODUCTION

Reliable decision-making in the exploration and production industry relies on building robust spatial models of the reservoir properties from seismic data. Deep learning (DL) models, such as convolutional neural networks (CNNs) and generative adversarial networks are being increasingly employed towards this end (Das et al., 2019; Das and Mukerji, 2020; Mosser et al., 2020, Pradhan and Mukerji, 2020a). DL models are particularly attractive because of their ability to learn highly non-linear functional relationships between the input and output variables by stacking together multiple non-linear neural network layers. However, a challenge for successful application of DL algorithms to subsurface problems is the lack of labeled training data. Optimization of DL model parameters requires multiple example pairs of the input variable or features and the output variable or labels. The primary challenge lies in the fact that reservoir properties (labels) are available at a sparse set of well locations in the modeling domain and seismic data (features) away from the well are unlabeled.

Several authors have proposed to address this challenge by generating training dataset synthetically. Many of these approaches rely on generating multiple earth models by stochastic perturbation of modeling parameters and forward modeling of the data variables. Wu et al. (2019, 2020) generate multiple 3D earth models for structural seismic interpretation with CNNs by stochastically generating folding and faulting structures in a layered earth model. Yang and Ma (2019) build velocity models for DL based velocity inversion by randomly assigning velocity values to the layers of a layered earth model. Many authors have proposed generating multiple synthetic earth property models by sampling using geostatistical algorithms (Caers, 2005). Das et al. (2019) created synthetic earth models for impedance inversion with CNNs using the sequential indicator simulation for simulation of reservoir facies and sequential Gaussian simulation (SGSIM) for simulation of petrophysical properties. Pradhan and Mukerji (2020a) employed object-based geostatistical models in conjunction with SGSIM to simulate synthetic training models for reservoir facies classification problem, while Mosser et al. (2020) use a generative adversarial network (GAN) as a geological prior for stochastic seismic waveform inversion.

While it is possible to train DL models with synthetically generated training data, using the model to make reliable predictions with real data comes with several pitfalls. One challenge derives from the fact that the synthetic training data might be statistically inconsistent with real data. Most DL models are designed to approximate the underlying statistical distribution exhibited in the training dataset. A DL model trained with a dataset that is statistically inconsistent with real data will lead to erroneous results during predictions with real data. Pradhan and Mukerji (2020b) elaborate on this issue within the context of estimation of sub-resolution reservoir properties from seismic data with unsupervised and supervised machine learning models. In this paper, we propose employing a falsification analysis to establish the consistency of the underlying prior geostatistical model, from which the training data is sampled, with real data. The approach of prior falsification analysis was first proposed by Scheidt et al. (2017) for performing Bayesian inference in reservoir forecasting applications. The novelty of this paper relates to the proposal of strategies appropriate for performing the falsification analysis with 3D partial angle-stack seismic data. A second pitfall of employing statistical distributions to generate synthetic training data derives from the fact that the DL model could potentially overfit to the statistical distribution. This could be an issue during predictions with real data since modeled synthetic data will always contain modeling imperfections and approximations, while the real data are noisy. In this paper, we present strategies for detecting this behavior during training of DL models and propose capturing the uncertainty arising due to modeling imperfections and data noise with geostatistical models. We demonstrate the efficacy of proposed strategies with the help of real case application of facies classification in an offshore deltaic reservoir with 3D partial angle-stack seismic data by deep 3D convolutional neural networks (CNNs).

In summary, the novel contributions of this paper are as follows.

1. We present strategies amenable to accomplishing prior falsification of synthetic training data for DL with 3D partial angle-stack seismic data.

2. We introduce the issue of overfitting of DL models to the synthetic training data distribution and propose geostatistical modeling of noise random variables as part of the synthetic training data generation. We propose strategies and practical examples of how the noise distribution may be used to prevent overfitting of the network to the synthetic training data distribution.

3. We present a real-world application of 3D reservoir facies classification with 3D seismic data and 3D CNNs from an offshore deltaic reservoir, demonstrating how the methods proposed in this paper may employed to obtain reliable predictions with real data using DL models trained on synthetic data.

The remainder of the paper is organized as follows. We begin with a theoretical description of the methods proposed by this paper. Specifically, we highlight the procedure of synthetic data generation, prior falsification, the issue of overfitting to the synthetic statistical distribution used to generate the training dataset and strategies for tackling it. This discussion on methodology is followed by the real case study from offshore Nile Delta, where we demonstrate the practical aspects and nuances of the methods proposed in this paper. We conclude by highlighting some of the limitations with the proposed approach and discuss ideas for future research directions.

METHODOLOGY

**Deep learning with convolutional neural networks**

In the following, reservoir facies, denoted by $\boldsymbol{h}$, are the target variables of interest or dependent variables. Seismic data, denoted by $\boldsymbol{d}$, are the independent variables. Given seismic data, the objective is to estimate corresponding reservoir facies variables. In this paper, we propose performing supervised learning with convolutional neural networks (CNNs). In CNNs, the relationship between the inputs and outputs is modeled through multiple non-linear hidden convolutional layers (Krizhevsky et al., 2012; Long et al., 2015; Goodfellow et al., 2016). We operate in discriminative learning settings, in which the input-output relationship is modeled as the conditional probability distribution $f_{\mathrm{CNN}}(\boldsymbol{h}|\boldsymbol{d};\theta)$ (Ng and Jordan, 2002). Here, $f_{\mathrm{CNN}}(\boldsymbol{h}|\boldsymbol{d};\theta)$ denotes the conditional probability distribution learned by the CNN model. A specific illustration of how the discriminative learning distribution is modeled with CNNs is provided in the application section. The distribution is parameterized by learnable parameters $\theta$, which are estimated with the help of a training dataset $\mathfrak{D} = \left\{\left(\boldsymbol{h}^i, \boldsymbol{d}^i\right)\right\}_{i=1}^n$. Parameter estimation is achieved by maximum likelihood estimation, where $\theta$ are estimated as

$$\theta = \arg\max_{\theta} \prod_{i=1}^{n} f_{\mathrm{CNN}}(\boldsymbol{h}^i|\boldsymbol{d}^i;\theta), \tag{1}$$

Goodfellow et al. (2016; chapter 5). The above equation states that we seek to find $\theta$ by maximizing the probability of the training examples under the supervised learning model distribution. Implicit in the above equation is the assumption that training examples are obtained independently of each other. As shown by Goodfellow et al. (2016), maximizing the likelihood of the training examples under the model distribution can be interpreted as minimizing a statistical distance between $f_{\text{CNN}}(\boldsymbol{h}|\boldsymbol{d};\theta)$ and the underlying true distribution $f_{\text{true}}(\boldsymbol{h}|\boldsymbol{d})$. Note that in the above formulation, $f_{\text{true}}(\boldsymbol{h}|\boldsymbol{d})$ does not need to be known, the only requirement being the availability of training samples from this distribution.

**Modeling synthetic training data for supervised learning: prior specification and sampling**

Obtaining repeated measurements of $\boldsymbol{h}$ and $\boldsymbol{d}$ for formation of training set $\left\{\left(\boldsymbol{h}^i, \boldsymbol{d}^i\right)\right\}_{i=1}^{n}$ is challenging, especially when the goal is to estimate facies on a 3D discretized grid of the subsurface. The vector $\boldsymbol{h}$ consists of the collection of facies variables at each voxel of the 3D grid. In typical exploration and development scenarios, subsurface property observations are available once at a subset of grid locations in the form of well measurements. 3D seismic data is also typically acquired once in the field, leading to a single realization $\boldsymbol{d}_{obs}$ for the data variable. To address the above challenges, many authors have proposed to create the training set synthetically as detailed in the Introduction section. In this paper, we employ an approach that facilitates systematic generation of training examples consistently with geological understanding of the subsurface system and geophysical principles of data generation. Specifically, training set examples are obtained by Monte-Carlo sampling from the probability distribution $f_{\text{synth}}(\boldsymbol{h}, \boldsymbol{d})$ specified a priori and decomposed as

$$f_{\text{synth}}(\boldsymbol{h}, \boldsymbol{d}) = f_{\text{geol}}(\boldsymbol{h}) f_{\text{geophys}}(\boldsymbol{d}|\boldsymbol{h}). \tag{2}$$

The distribution $f_{\text{geol}}(\boldsymbol{h})$ is used to model prior knowledge and uncertainty on $\boldsymbol{h}$ as derived from geological understanding and reasoning about the subsurface variability of $\boldsymbol{h}$. Geostatistical modeling methods (Caers, 2005; Pyrcz and Deutsch, 2014) are commonly employed for probabilistic modeling of subsurface heterogeneities encountered in a wide range of geological settings. The conditional distribution $f_{\text{geophys}}(\boldsymbol{d}|\boldsymbol{h})$ is used to model the physical relationship between $\boldsymbol{h}$ and $\boldsymbol{d}$ and associated uncertainties, and is typically specified using a deterministic geophysical forward model $g_{\text{geophys}}(.)$ and noise random variable $\boldsymbol{\epsilon}$

$$\boldsymbol{d} = g_{\text{geophys}}(\boldsymbol{h}) + \boldsymbol{\epsilon}. \tag{3}$$

Training set $\left\{\left(\boldsymbol{h}^i, \boldsymbol{d}^i\right)\right\}_{i=1}^{n}$ is created by Monte-Carlo sampling of $\boldsymbol{h}^i$s from $f_{\text{geol}}(\boldsymbol{h})$ and employing equation 3 to generate corresponding $\boldsymbol{d}^i$s. Thus, the prior distribution $f_{\text{synth}}(\boldsymbol{h}, \boldsymbol{d})$ is used as a proxy for the unknown and difficult-to-sample-from true distribution $f_{\text{true}}(\boldsymbol{h}, \boldsymbol{d})$.

**Ensuring consistency between modeled and real data: prior falsification**

The practice of modeling prior uncertainty through probability distributions is common in parameter estimation by Bayesian inversion methods (Tarantola, 2005) and it is worthwhile highlighting some of the similarities and resulting caveats shared with these methods. In the Bayesian formulation, given observed data $\boldsymbol{d}_{obs}$, the goal is to estimate or sample from the posterior distribution $f(\boldsymbol{h}|\boldsymbol{d}_{obs})$. Using Bayes' rule, the posterior can be expressed in terms of the joint probability distribution over $\boldsymbol{h}$ and $\boldsymbol{d}$ as

$$f(\boldsymbol{h}|\boldsymbol{d}_{obs}) = \frac{f(\boldsymbol{d}_{obs}, \boldsymbol{h})}{f(\boldsymbol{d}_{obs})}, \tag{4}$$

where $f(\boldsymbol{d}_{obs}, \boldsymbol{h})$ is factored as shown in equation [2]. Given that a priori specification of $f_{\text{geol}}(\boldsymbol{h})$ and $f_{\text{geophys}}(\boldsymbol{d}|\boldsymbol{h})$ requires making subjective modeling choices, a caveat is that the prior distribution might be inconsistent with $\boldsymbol{d}_{obs}$. Mathematically, such an inconsistent prior model would make $f_{\text{geophys}}(\boldsymbol{d}_{obs}|\boldsymbol{h})$ evaluate to negligible values given any sample $\boldsymbol{h}$ from $f_{\text{geol}}(\boldsymbol{h})$. In our case, supervised learning model trained on samples from inconsistent $f_{\text{synth}}(\boldsymbol{h}, \boldsymbol{d})$ will potentially lead to biased predictions with $\boldsymbol{d}_{obs}$. Similar to Bayesian inverse problems, it is imperative to employ explicit measures for establishing consistency of $f_{\text{synth}}(\boldsymbol{h}, \boldsymbol{d})$ prior to training of the supervised learning model.

To this end, we propose performing a falsification analysis of the prior model (Scheidt et al., 2017). A quantitative comparison is made between modeled data $\{\boldsymbol{d}^i\}_{i=1}^n$ and real data $\boldsymbol{d}_{obs}$ to determine whether they belong to the same statistical population. In the event that $f_{\text{synth}}(\boldsymbol{h}, \boldsymbol{d})$ deviates significantly from $f_{\text{true}}(\boldsymbol{h}, \boldsymbol{d})$, $\boldsymbol{d}_{obs}$ can be expected to feature as an outlier with respect to the training set. In that case, the prior distribution will be falsified, necessitating modifications to the $f_{\text{synth}}(\boldsymbol{h}, \boldsymbol{d})$, such as broadening the range of parameters or changing the geological conceptual model. Typically, an outlier detection algorithm is utilized to make the above determination. For high dimensional datasets, a commonly employed algorithm is the Mahalanobis distance-based outlier detection (Rousseeuw and Van Zomeren, 1990). To identify outliers in a given set of $m$-dimensional samples $\{\boldsymbol{d}^i \in \mathbb{R}^m; i = 1, .., n\}$, the Mahalanobis distance is computed as

$$d^M = \sqrt{(\boldsymbol{d} - \hat{\boldsymbol{\mu}})^T \hat{\boldsymbol{\Sigma}}^{-1}(\boldsymbol{d} - \hat{\boldsymbol{\mu}})}, \qquad (5)$$

where $\hat{\boldsymbol{\mu}}$ and $\hat{\boldsymbol{\Sigma}}$ are robust estimates of the data mean and covariance, estimated using the minimum covariance determinant estimator proposed by Rousseeuw and Van Driessen, 1999. The robust estimation technique makes the mean and covariance estimation less sensitive to the presence of outliers in the data. Note that assigning a constant value to $d^M$ is equivalent to defining an ellipsoid in $\mathbb{R}^m$ centered at $\hat{\boldsymbol{\mu}}$. A threshold $\tau$ is assigned such that $\boldsymbol{d}_i$s located outside the ellipsoid defined by $\tau$ are deemed as outliers. The threshold is typically assigned as $\tau = \sqrt{\chi^2_{m,0.975}}$, where $\chi^2_{m,0.975}$ is the 97.5% quantile of the chi-square distribution with $m$ degrees of freedom. The choice for the threshold can be motivated from the fact that if the underlying data distribution is a multivariate Gaussian, then the probability that $d^{M^2} > \chi^2_{m,0.975}$ is $(1 - 0.975)$. Note that there is a finite probability that some samples from the prior distribution, in a large training set, may have $d^M$ greater than $\tau$.

In many cases, it might be desirable to perform the falsification analysis with respect to specific summary statistics extracted from the data. For instance, Scheidt et al. (2015) discuss how a local trace-by-trace comparison of seismic datasets might not be effective in falsifying prior geological distributions, as opposed to a comparison based on global features or patterns in the data which may identify if the synthetic prior distribution is simulating patterns inconsistent with the true geology. For falsification of the prior model with seismic data in the application section, we adapt the methodology proposed by Scheidt et al. (2015), in which a global patterns-based comparison is performed using discrete wavelet transform. We provide additional details of the methodology in the application section.

**Overfitting to synthetic training distribution and remedial measures**

Overfitting to the training dataset is a common issue in supervised learning applications (Bishop, 2006) and occurs when the trained $f_{\text{CNN}}(\boldsymbol{h}|\boldsymbol{d};\theta)$ exhibits low prediction errors on the training dataset but fails to reproduce similar prediction performance on examples not included in the training set. This happens when parameters $\theta$ are optimized to fit noisy patterns and features present in the training dataset, resulting in a model limited in its generalization power to previously unseen examples. The technique of cross-validation is commonly employed to identify overfitting, in which the training dataset is split into training and validation sets. The latter, unlike the former, is kept hidden during optimization of the network parameters. A model that is overfitting will exhibit high prediction accuracy or low prediction error on the training set and vice versa for the validation set.

Specific to our proposed methodology is the problem of overfitting to the synthetic training data distribution. Note that the problem pertains to overfitting to the training data distribution and not to a specific training set as discussed above. Take for instance the CNN model $f_{\text{CNN}}(\boldsymbol{h}|\boldsymbol{d};\theta)$ optimized using training set $\mathfrak{D}_{\text{train}} = \left\{\left(\boldsymbol{h}^i_{\text{train}}, \boldsymbol{d}^i_{\text{train}}\right)\right\}_{i=1}^{n_{\text{train}}}$, constituting of samples from $f_{\text{synth}}(\boldsymbol{h}, \boldsymbol{d})$. Consider two different validation sets for cross-validation purposes. The first set, $\mathfrak{D}_{\text{val}} = \left\{\left(\boldsymbol{h}^i_{\text{val}}, \boldsymbol{d}^i_{\text{val}}\right)\right\}_{i=1}^{n_{\text{val}}}$, constitutes of examples randomly sampled from $f_{\text{synth}}(\boldsymbol{h}, \boldsymbol{d})$, thus distinct from $\mathfrak{D}_{\text{train}}$. The second set $\mathfrak{D}_{\text{real}} = \{(\mathbb{S}[\boldsymbol{h}_{\text{true}}], \boldsymbol{d}_{obs})\}$ consists of the real data and a subset of labels from the true facies model $\boldsymbol{h}_{\text{true}}$. Here, $\mathbb{S}[.]$ denotes the operator extracting the true facies labels at a subset of locations such as wells. In other words, $\mathfrak{D}_{\text{real}}$ is used to validate model predictions using real data. If the CNN demonstrates high prediction accuracy simultaneously on $\mathfrak{D}_{\text{train}}$ and $\mathfrak{D}_{\text{val}}$, but relatively poor performance on $\mathfrak{D}_{\text{real}}$, a possible reason could be that the network is overfitting to $f_{\text{synth}}(\boldsymbol{h}, \boldsymbol{d})$. Note that while prior falsification analysis identifies statistically significant deviations of $f_{\text{synth}}(\boldsymbol{h}, \boldsymbol{d})$ from $f_{\text{true}}(\boldsymbol{h}, \boldsymbol{d})$, minute deviations are expected to exist due to modeling imperfections and data noise. A difference in cross-validation performance on $\mathfrak{D}_{\text{val}}$ and $\mathfrak{D}_{\text{real}}$ might be effectuated if $f_{\text{CNN}}(\boldsymbol{h}|\boldsymbol{d}; \theta)$ overfits to these modeling noise during training and $\boldsymbol{d}_{obs}$ contains data noise during prediction. In the next section, we present empirical observations of overfitting to the training distribution using an example of a deep learning model trained on synthetic data.

Different kinds of regularization techniques have been proposed in the machine learning literature (Bishop, 2006) to tackle overfitting. We adapt two regularization strategies that were found to be effective in our case. The first technique is based on early stopping of the training based on cross-validation with $\mathfrak{D}_{\text{real}}$. This entails evaluating the network's performance on $\mathfrak{D}_{\text{real}}$ after each epoch of the training process and retaining parameters $\theta$ estimated during the best performing epoch. The second strategy consists of corrupting the training samples with noise. Bishop (1995) has shown that adding random noise to training examples is equivalent to performing Tikhonov regularization (Tikhonov and Arsenin, 1977). To prevent overfitting to the training distribution, we propose adding noise to the training samples such as to compensate for modeling imperfections and data noise. This can be accomplished naturally within our framework through the noise random variable $\boldsymbol{\epsilon}$ introduced previously. We adopt a two-step strategy to choose a pertinent probability distribution for $\boldsymbol{\epsilon}$. We first employ the prior falsification analysis to establish the general form of the distribution, for instance spatially uncorrelated distribution vs. spatially correlated distribution. The noise-to-signal level that is effective in preventing overfitting will generally vary depending on the specific machine learning model under consideration and its architectural characteristics. We propose determining the signal-noise ratio based on cross-validation with $\mathcal{D}_{real}$. Additional implementational details are presented in the next section.

## REAL CASE APPLICATION

In this section, we present a real case study of reservoir facies classification with 3D seismic data. Specifically, we show how our proposed approach may be used to generate a large number of training examples, consistent with prior geological knowledge and geophysical forward modeling, and may be used in the supervised classification problem without overfitting. The area of interest is located in offshore Nile delta, with a gas producing reservoir as part of slope-channel system of Plio-Pleistocene age presently at a depth of 2100 m. Aleardi and Ciabarri (2017), and Aleardi et al. (2018) have previously performed studies on rock physics modeling and probabilistic seismic petrophysical inversion respectively in this field. Our analysis will be performed on an area with a spatial extent of 5 kms along $x$ and $y$ directions. The thickness of the reservoir zone varies spatially as shown in [Figure 1](#) with an average thickness of 250 meters. The zone of interest was discretized into 100, 100 and 250 cells along $x$, $y$ and depth dimensions, respectively. Shaly over and under-burden zones, each 50 meters thick, were added to the model as shown in [Figure 1](#). Details of the dataset are presented below.

1. Log data from six wells in the study area with petrophysical facies interpretations of the following facies were available: channels, splays, levees, thin beds and shales. Interpreted facies were chosen to be consistent with previous seismic geomorphological analyses of the reservoir zone (Cross et al., 2009), which have identified the presence of amalgamated, sinuous and leveed sand channels along with thinly bedded sandstone and limestone facies. Compressional sonic, shear sonic, density and water saturation logs were also available. Four wells, termed wells 1-4, will be used in specification of the prior distribution for sampling of the training datasets. Wells 5-6 are kept blind during the prior building process.

2. Seismic data consisted of 3D time migrated post stack volume, near stack volume obtained by stacking $2^0$, $7^0$ and $12^0$ angle gathers and far stack volumes obtained from $21^0$, $26^0$ and $30^0$ angle gathers. The data has a time sampling of 4 ms and a dominant frequency of 18 Hz. The seismic data grid in the zone of interest was discretized into 100, 100 and 60 cells along $x$, $y$ and time dimensions respectively. To avoid aliasing effects, time discretization of 60 cells was chosen to retain the original sampling rate of 4 ms, the original sampling rate of the data, at the location with maximum reservoir thickness. Figure 2 shows the post, near and far stack seismic data on the seismic grid.

In this study, we employ deep 3D CNNs to estimate reservoir facies from seismic data. As discussed previously, we seek to learn the relationship $f_{\text{CNN}}(\boldsymbol{h}|\boldsymbol{d}; \theta)$. Here $\boldsymbol{h}$ and $\boldsymbol{d}$ are random vectors consisting of random variables representing each voxel in the model and data grids respectively (Figure 1 and Figure 2).

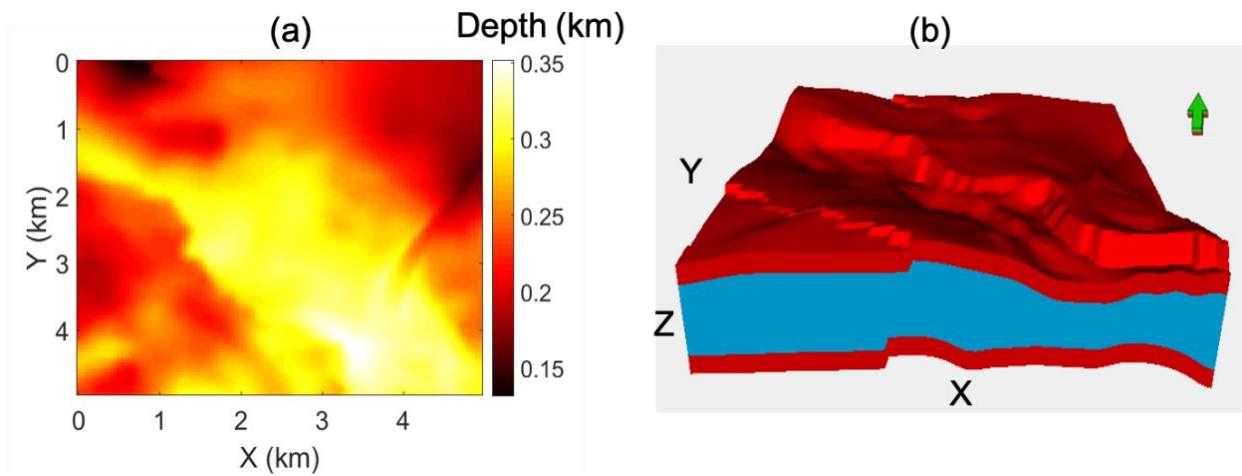

Figure 1: (a) Depth thickness map of the reservoir zone in the area of interest. (b) The reservoir model grid used in this study. The zone of interest is shown in blue, and the over-burden and under-burden zones are shown in red.

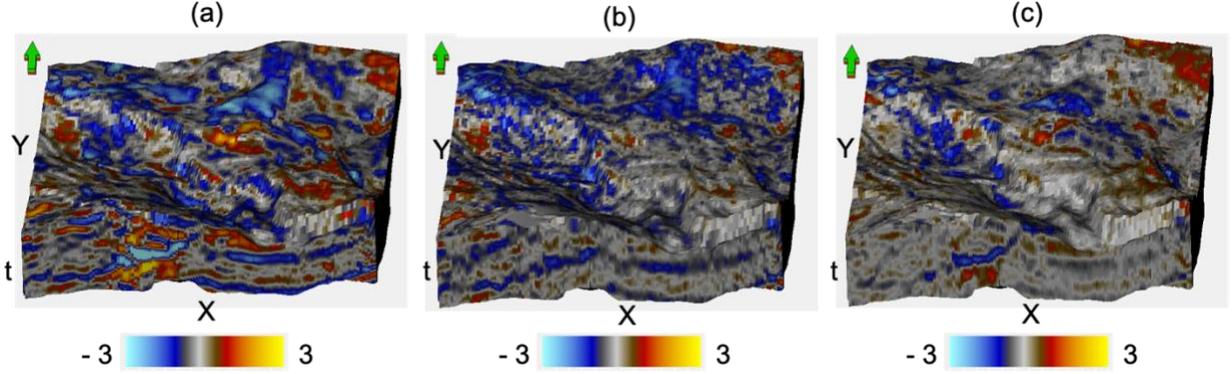

Figure 2: Field (a) post-stack, (b) near stack and (c) far stack seismic volumes sampled into the data grid.

**Specification of the prior uncertainty model**

Training examples will be generated by sampling the synthetic probability distribution $f_{\text{synth}}(\boldsymbol{h}, \boldsymbol{m}, \boldsymbol{d})$ decomposed as

$$f_{\text{synth}}(\boldsymbol{h}, \boldsymbol{m}, \boldsymbol{d}) \approx f_{\text{WP}}(\boldsymbol{d}|\boldsymbol{m}) f_{\text{RP}}(\boldsymbol{m}|\boldsymbol{h}) f_{\text{geol}}(\boldsymbol{h}). \qquad (6)$$

Here, $\boldsymbol{m} = [\boldsymbol{m}_{V_P}, \boldsymbol{m}_{V_S}, \boldsymbol{m}_{\rho_b}]^T$ is the random vector for rock elastic properties such as P-wave velocity $V_P$, S-wave velocity $V_S$ and bulk density $\rho_b$. Distributions $f_{\text{RP}}(.)$ and $f_{\text{WP}}(.)$ denote the rock physics and wave propagation forward model distributions respectively, while $f_{\text{geol}}(.)$ models the prior geologic uncertainty. We describe below how we specify each of these distributions below.

*Model for prior geological uncertainty*

As discussed by Cross et al. (2009), the stratigraphic elements of the reservoir system exhibit channelized geometries and spatial patterns characteristic of deep-water slope channel systems. It is thus desirable that $f_{\text{geol}}(\boldsymbol{h})$ is able to capture and simulate the complex geological patterns consistently with prior geological knowledge. To this end, we use Boolean or object-based geostatistical model, which can be used to simulate geologically realistic earth models by stochastically dropping objects representing facies into the modeling grid (Pyrcz and Deutsch, 2014). Five facies are considered as objects in the simulations: channels, splays, levees, thin beds and shales. Shale is the background facies. Leveed channels, splays and thin bed facies are modeled with sinusoids, fan lobes and ellipsoidal objects respectively. The prior $f_{\text{geol}}(\boldsymbol{h})$ is specified through probability distributions on the uncertain simulation parameters, such the volumetric proportion and parameters controlling the shape and geometry of each facies object, as shown in Table 1. These probability distributions were specified primarily based on the geomorphological analysis of the reservoir architecture performed by Cross et al. (2009). For instance, channel bodies were determined to exhibit a maximum thickness of 30 meters in the above study; consequently, we assigned the prior on channel thickness parameter as a triangular distribution $\mathcal{T}(2\ m, 15\ m, 30\ m)$. We used Petrel commercial software to generate an initial set of 500 unconditional object simulations of the five facies objects under the prior model (Figure 3). These realizations will be used in the prior falsification analysis detailed below. Note that no hard data is used during the simulation of the facies realizations. The simulated channel objects are subsequently classified into gas and brine saturated channels. Using water saturation logs available at the wells, we assumed the prior for gas-water contact (GWC) depth to follow a triangular distribution with lower and upper limits of 2415 and 2450 meters respectively and 2430 meters as mode. For every facies realization, a sample of the GWC depth value is obtained from its prior and used to assign the fluid saturation scenarios in the channel objects.

Table 1: Prior distributions on the parameters for facies geo-objects. $\mathcal{N}(\mu,\sigma)$: Normal distribution with mean $\mu$ and standard deviation $\sigma$. $\mathcal{T}(a,c,b)$: triangular distribution with upper and lower limits of $a$ and $b$ respectively, and mode $c$. $U(a,b)$: uniform distribution with upper and lower limits of $a$ and $b$.

| Geo-object | Parameters | Distributions |
|---|---|---|
| Channel | Global proportion | $\mathcal{N}(27\%, 6\%)$ |
| Channel | Amplitude | $\mathcal{T}(300\ m, 400\ m, 500\ m)$ |
| Channel | Wavelength | $\mathcal{T}(350\ m, 600\ m, 800\ m)$ |
| Channel | Width | $\mathcal{T}(100\ m, 250\ m, 450\ m)$ |
| Channel | Thickness | $\mathcal{T}(2\ m, 15\ m, 30\ m)$ |
| Channel | Orientation | $\mathcal{T}(240^0, 305^0, 310^0)$ |
| Levee | Width (fraction of channel width) | $\mathcal{T}(0.1,\ 0.3,\ 0.8)$ |
| Levee | Thickness (fraction of channel width) | $\mathcal{T}(0.3, 0.6,\ 0.9)$ |
| Splay | Global proportion (fraction of channel proportions) | $U(10\%, 50\%)$ |
| Splay | Minor width | $\mathcal{T}(180\ m,\ 350\ m, 700\ m)$ |
| Splay | Major to minor width ratio | $\mathcal{T}(0.7,\ 1.2,\ 2.2)$ |
| Splay | Thickness | $\mathcal{T}(2\ m, 4\ m, 9\ m)$ |
| Thin beds | Global proportion | $\mathcal{N}(12\%, 4\%)$ |
| Thin beds | Minor width | $\mathcal{T}(600\ m,\ 800\ m, 4000\ m)$ |
| Thin beds | Major to minor width ratio | $\mathcal{T}(0.8,\ 1,\ 1.2)$ |
| Thin beds | Thickness | $\mathcal{T}(1\ m, 1.5\ m, 2\ m)$ |
| Thin beds | Orientation | $\mathcal{T}(300^0, 305^0, 310^0)$ |

*Rock physics modeling*

Random variables for $V_P, V_S$ and $\rho_b$ after conditioning to facies are assumed to be linearly correlated and distributed according to multivariate Gaussian distributions. Specifically, we specify that $f_{\mathrm{RP}}(\varphi(\boldsymbol{m}_l)|\boldsymbol{h}=\boldsymbol{j})\sim\mathcal{N}\left(\boldsymbol{\mu}_l^j, \boldsymbol{\Sigma}_l^j\right)$, with mean vector $\boldsymbol{\mu}_l^j$ and covariance matrix $\boldsymbol{\Sigma}_l^j$. We use subscript $l$ to denote each of the three elastic properties under consideration, while superscript $j$ denotes the facies category with $j$ taking values from $\{1, .., k\}$. Since earth properties cannot be expected to be normally distributed in general, the normal distribution is assigned after applying a normal-score transform $\varphi(.)$ (Deutsch and Journel, 1998) to the original variable. Sonic, bulk density and legacy petrophysical facies logs were used to estimate the global facies-conditional distributions for each property before normal score transformation. In [Figure 4](#), we show the bivariate distributions for P-impedance and $V_P - V_S$ ratio. The spatial covariance matrices of the Gaussian distributions were specified using parametric variogram models (Goovaerts, 1997), shown in [Table 2](#). Correlation ranges along the vertical direction (minor range) were estimated by fitting to the experimental variogram calculated at the wells. Note that it is challenging to estimate the horizontal (major and medium) ranges directly from well data due to absence of horizontal continuity in the data. The ratio of the vertical to horizontal variogram range of the post-stack seismic data was used as a guide to roughly assign the horizontal ranges shown in [Table 2](#). For any $\boldsymbol{h}$ sampled from $f_{\text{geol}}(\boldsymbol{h})$, a corresponding realization from $f_{\text{RP}}(\boldsymbol{m}|\boldsymbol{h})$ is generated using sequential simulation algorithms. To elaborate, facies conditional realizations of $V_P$ were first generated by sequential Gaussian simulation (SGSIM) (Deutsch and Journel, 1998). Correlation coefficients of $V_S$ and $\rho_b$ with $V_P$ were estimated using well logs as shown in [Table 3](#). Given a realization of $V_P$, corresponding realizations of $V_S$ and $\rho_b$ for each facies are generated by sequential Gaussian co-simulation (COSGSIM), with correlations imposed by the Markov type-1 model (Goovaerts, 1997). Simulations of elastic properties were performed for all facies. Two realizations each of $V_P, V_S$ and $\rho_b$ obtained by sequential simulations are shown in [Figure 3](#). Prior to forward modeling of the seismic data, the generated realizations were upscaled by a running Backus averaging with a vertical window length of 13.5 meters, roughly corresponding to 1/10 of the seismic wavelength. (Avseth et al., 2005).

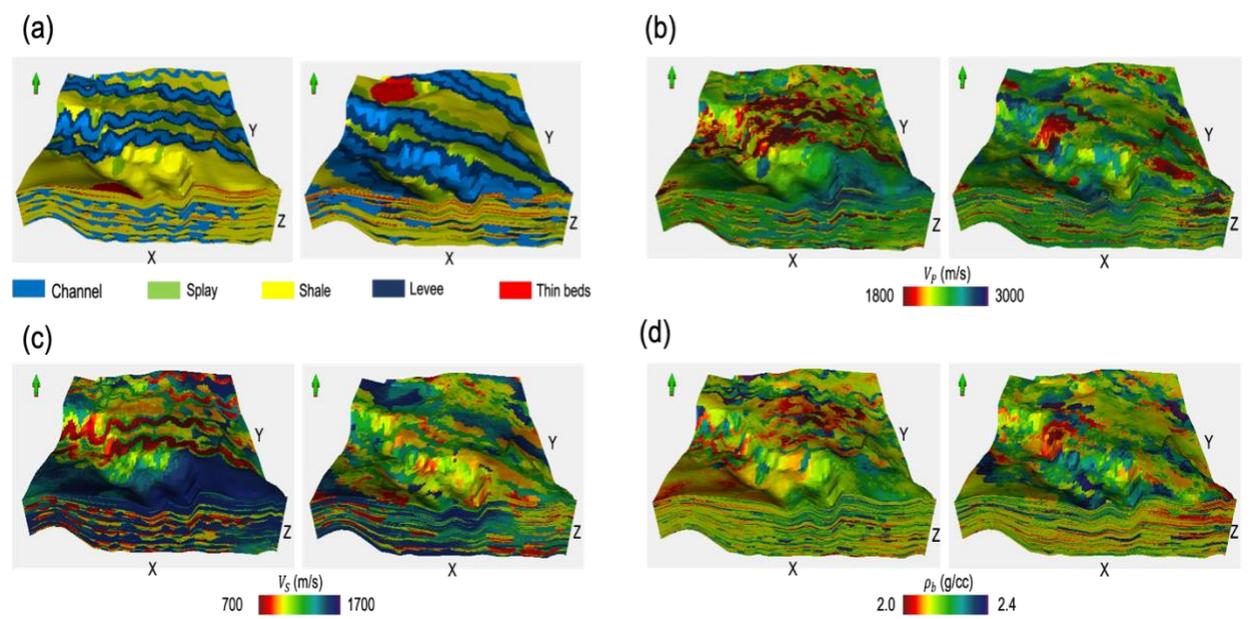

Figure 3: Two unconditional object model realizations of facies (a), $V_P$ (b), $V_S$ (c) and $\rho_b$ (d) sampled from the prior distribution.

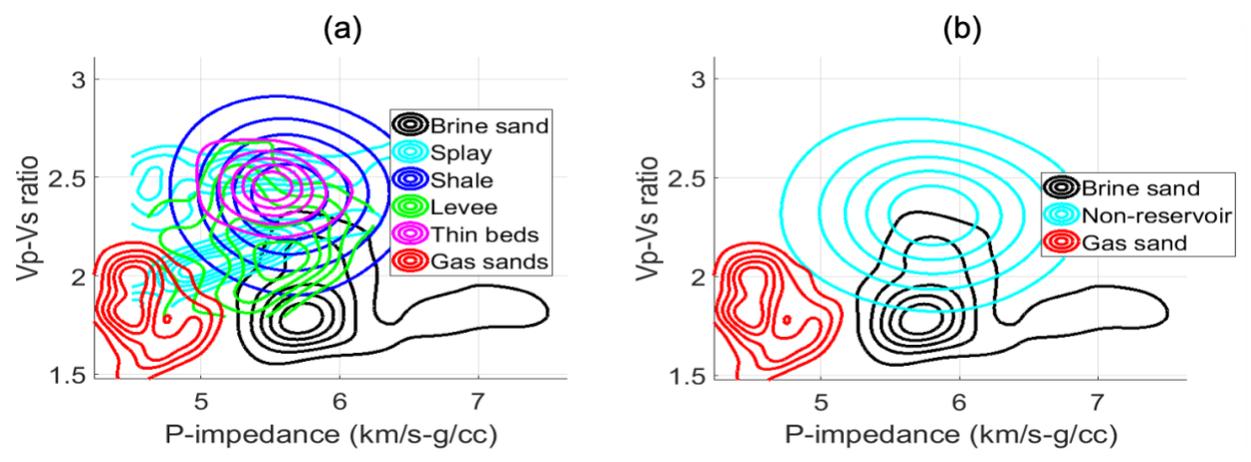

Figure 4: (a) Contour plots of the bivariate probability distributions for P-impedance and $V_P$-$V_S$ ratio for all facies. (b) Corresponding plots after lumping together all non-reservoir facies.

Table 2: Variogram parameters for facies conditional elastic properties used in the real case study.

| Facies | Property | Type | Major/medium/minor ranges (meters) |
|---|---|---|---|
| Channels | Density | Spherical | 1500/1500/10 |
| Channels | P-velocity | Spherical | 2000/2000/10 |
| Channels | S-velocity | Spherical | 2000/2000/10 |
| Splays | Density | Spherical | 1500/1500/10 |
| Splays | P-velocity | Spherical | 2000/2000/10 |
| Splays | S-velocity | Spherical | 2000/2000/10 |
| Shale | Density | Exponential | 2000/2000/5 |
| Shale | P-velocity | Exponential | 2500/2500/110 |
| Shale | S-velocity | Exponential | 2500/2500/110 |
| Levee | Density | Spherical | 2000/2000/10 |
| Levee | P-velocity | Spherical | 2000/2000/40 |
| Levee | S-velocity | Spherical | 2000/2000/40 |
| Thin beds | Density | Spherical | 1000/1000/5 |
| Thin beds | P-velocity | Spherical | 1200/1200/20 |
| Thin beds | S-velocity | Spherical | 1200/1200/20 |

Table 3: Correlation coefficients of the S-wave velocity and bulk density with P-wave velocity for different facies as estimated using wells logs.

| Facies | Correlation coefficient between $V_P$ and $V_S$ | Correlation coefficient between $V_P$ and $\rho_b$ |
|---|---|---|
| Brine saturated channels | 0.5 | 0.56 |
| Splays | 0.34 | 0.46 |
| Shales | 0.27 | 0.57 |
| Levees | 0.34 | 0.59 |
| Thin beds | 0.28 | 0.72 |
| Gas saturated channels | 0.24 | 0.77 |

*Seismic forward modeling*

Distribution $f_{WP}(\boldsymbol{d}|\boldsymbol{m})$ is specified as $\boldsymbol{d} = g_{WP}(\boldsymbol{m}) + \boldsymbol{\epsilon}$, where $g_{WP}(.)$ is the deterministic single-scattering forward model with the exact nonlinear Zoeppritz equation (Aki and Richards, 1980), and $\boldsymbol{\epsilon}$ is the random vector representing modeling imperfections and data noise. We considered two different approaches for estimation of wavelets to be used in forward modeling of the seismic data. In the first approach, wavelets for modeling of post, near and far stack seismic data were extracted at well 1 (Figure 5) by the spectral coherency matching technique proposed by Walden and White (1998). An advantage of this method is that it generates good seismic-to-well ties since wavelets are extracted by explicitly matching the forward modeled and observed traces at the well location. As shown in Figure 6, the resulting seismic-to-well ties exhibit high correlation coefficients of 87%, 81% and 87% for post, near and far stacks respectively. Even though optimal ties were obtained at the well-location, we found that the amplitude spectrum of these wavelets did not match that of the real seismic data (Figure 7). Hence, we generated a second set of wavelets (Figure 5) by the statistical wavelet extraction technique (Yi et al., 2013), which extracts wavelets by matching the amplitude spectrum of the seismic data. As shown in Figure 7, the spectrum of the statistical wavelet exhibits a significantly better match with the seismic data as compared to the former wavelet. Note, however, that the statistical wavelet extraction technique does not provide any information about the phase of the wavelets. We estimated the wavelet phase by comparing the correlation coefficients of the seismic-well ties at well 1. Zero-phase wavelets yielded the best seismic-well ties, with correlation coefficients of 70%, 67% and 81% for post, near and far stacks respectively (Figure 8). Prior falsification analysis, presented in the next section, indicated that the statistical wavelets are consistent with the acquired seismic data as compared to the Walden and White (1998) wavelets. Hence, the former wavelets were used in forward model $g_{WP}(.)$ to generate seismic angle gathers shown in Figure 9. Since it is difficult to estimate the true amplitude of the seismic wavelet, we performed global normalization of the modeled data realizations to bring them to the same scale as the real data. For instance, for post stack data, we calculated a single global scaling coefficient as the $\frac{\sqrt{Var[\boldsymbol{d}_{obs}^{post}]}}{\frac{1}{n}\sum_{i=1}^{n}\sqrt{Var[\boldsymbol{d}_{i}^{post}]}}$, where $Var[.]$ denotes the variance operator, $\boldsymbol{d}_{obs}^{post}$ denotes the real post stack data and $\boldsymbol{d}_{i}^{post}$ denotes corresponding prior realization. Each modeled post stack prior realization is subsequently multiplied by above global scaling coefficient. Similar scaling is performed for the near and far stack modeled realizations.

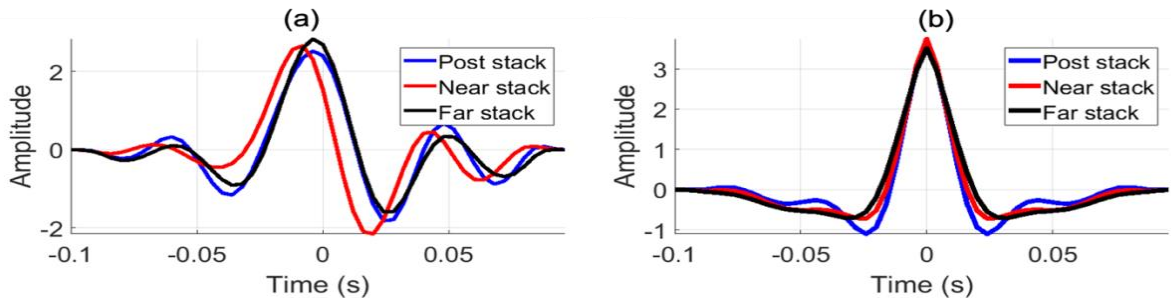

Figure 5: (a) Wavelets extracted by the coherency matching technique of Walden and White (1998). (b) Wavelets extracted by the statistical wavelet extraction technique.

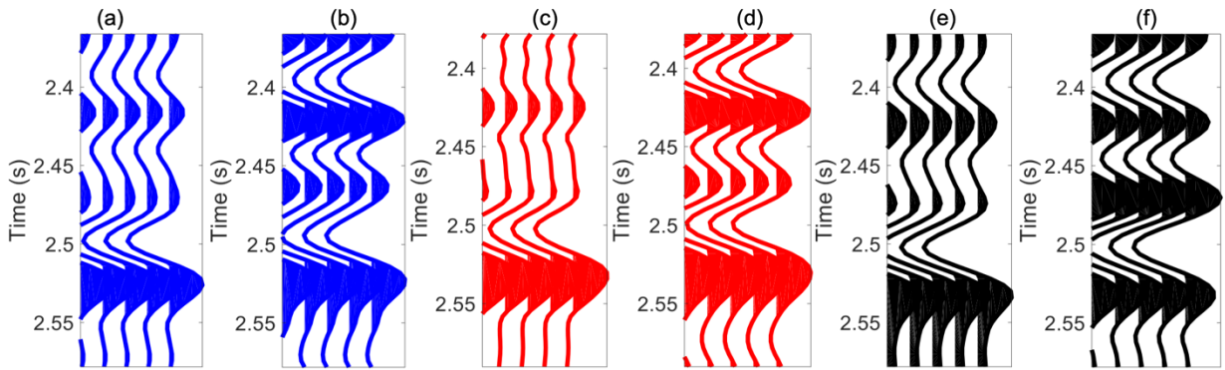

Figure 6: Seismic-well ties at well 1 obtained with Walden and White (1998) wavelets. Post-stack observed (a) and synthetic (b) seismic traces. Corresponding plots are shown in same sequence for near-stack (c-d) and far-stack traces (e-f). In every plot, a single trace is repeated five times for visual convenience.

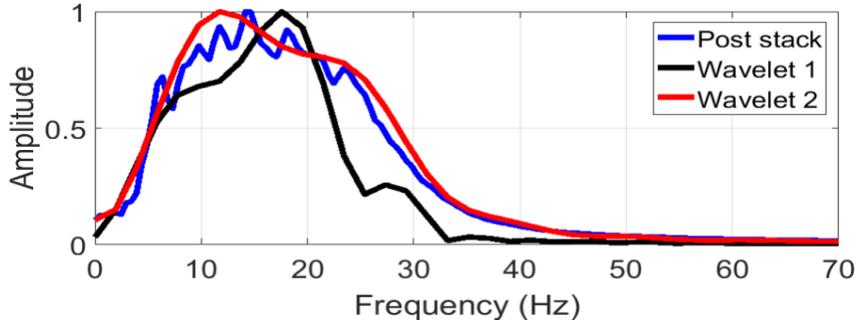

Figure 7: Comparison of amplitude spectrum of real post stack data with the spectra of the two post-stack wavelets analyzed in this chapter. Wavelet 1 refers to Walden and White (1998) wavelet and wavelet 2 refers to the statistical wavelet.

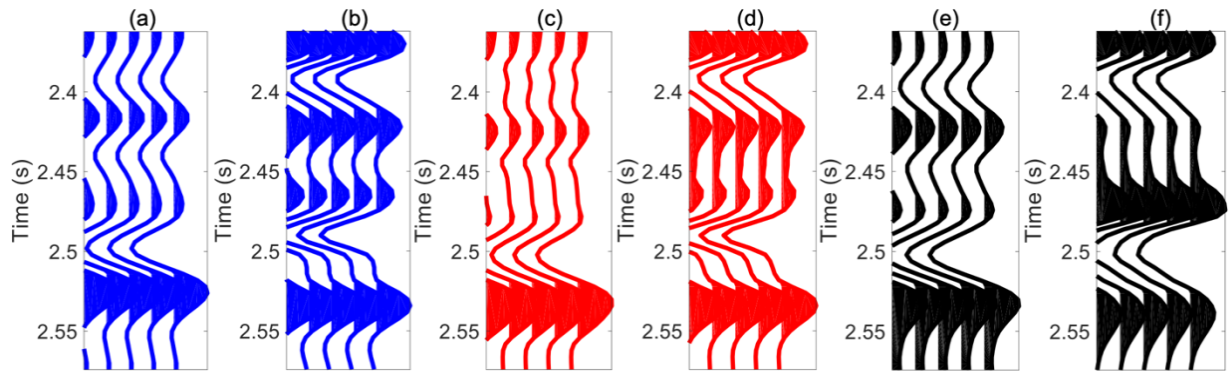

Figure 8: Seismic-well ties at well 1 obtained with statistical wavelets. Post-stack observed (a) and synthetic (b) seismic traces. Corresponding plots are shown in same sequence for near-stack (c-d) and far-stack traces (e-f). In every plot, a single trace is repeated five times for visual convenience.

*Noise modeling*

Data and forward-modeling noise is modeled with the random vector $\boldsymbol{\epsilon} = [\boldsymbol{\epsilon}^{post}, \boldsymbol{\epsilon}^{near}, \boldsymbol{\epsilon}^{far}]^T$. Random vectors $\boldsymbol{\epsilon}^{post}, \boldsymbol{\epsilon}^{near}, \boldsymbol{\epsilon}^{far}$ are assumed to be independent and identically distributed. We consider following three types of probability distributions for the noise vectors. Prior falsification analysis is subsequently employed to select the noise distribution consistent with $\boldsymbol{d}_{obs}$.

1. Noise distribution 1: Random variables at all spatial locations are assumed to be independent and identically distributed according to the univariate zero-mean Gaussian distribution. The variance is assigned based on the signal-noise ratio as discussed below.

2. Noise distribution 2: Each random vector is distributed according to the zero mean multivariate Gaussian distribution with finite correlation along the vertical direction and negligible correlations along the horizontal directions. Our motivation for this choice derives from the fact that residuals between the forward modeled and observed traces at well 1 (Figure 10) have finite correlations along the vertical direction. Specifically, we calculated the experimental variogram ranges along the vertical direction for the residuals to be approximately 70 ms across the three seismic stacks. We specified covariance matrices of the noise random vectors with spherical variogram models with minor range of 70 ms. The major ranges were assumed to be negligible in this scenario.

3. Noise distribution 3: In the above scenario, we assume the zero mean multivariate Gaussian noise random vectors to have finite horizontal correlations in addition to vertical correlations. This scenario is considered to model horizontally correlated noise that may potentially arise from modeling imperfections in the geological or geophysical forward model. To account for both short and long-range correlations, we assumed that the major and medium variogram ranges are random variables, distributed as $U(50\ meters, 2000\ meters)$. $U(a, b)$ denotes the uniform distribution with upper and lower limits of $a$ and $b$. The upper limit of 2000 meters was chosen to roughly correspond to the variogram ranges of the elastic properties (Table 2).

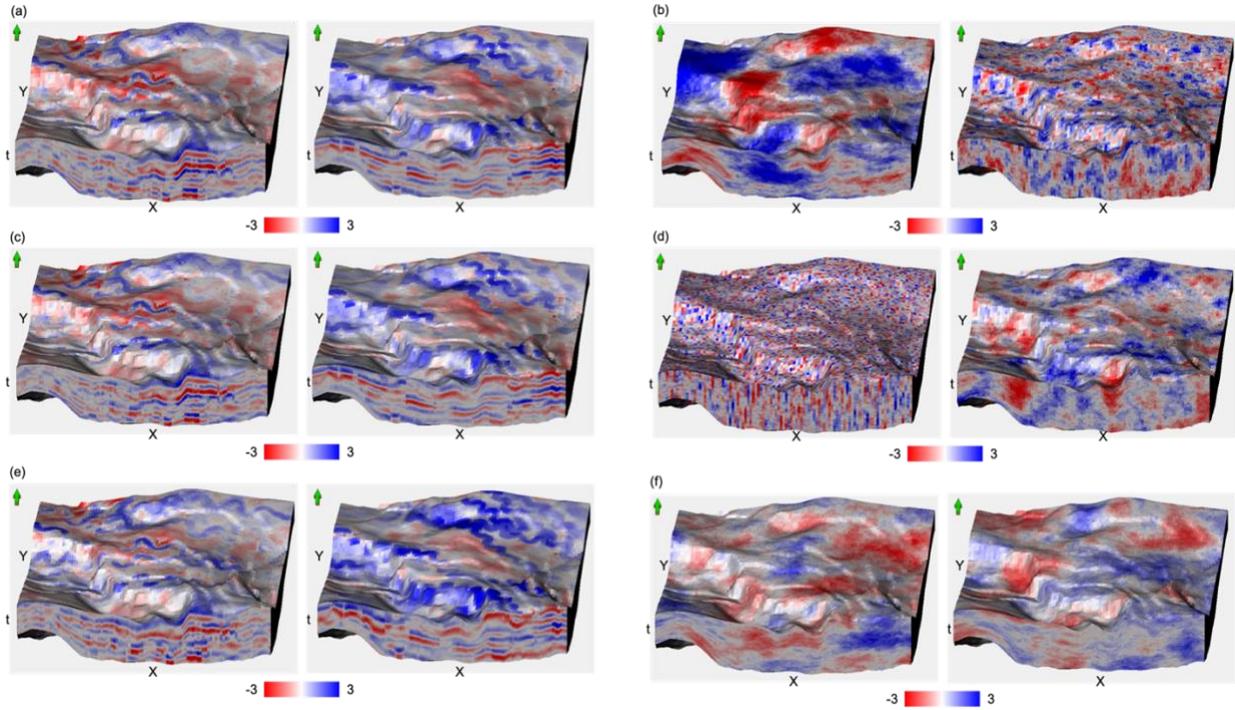

Figure 9: Forward modeled post-stack (a), near-stack (b) and far-stack (c) seismic data and corresponding noise volumes (b, d, f) for the realizations shown in Figure 3.

The global variance for the cases described above is assigned using the signal-noise ratio, specified as $N/S = \dfrac{Var[noise]}{Var[signal]}$, with $Var[.]$ denoting the variance operator. The true $N/S$ is unknown. While it is possible to derive an estimate at well 1, it might not be representative of the $N/S$ away from the well locations. Hence, we treat $Var[noise]$ as a random variable with tunable upper and lower bounds. Specifically, we take $Var[\boldsymbol{\epsilon}^{post}] \sim U(lb_{noise}\%, ub_{noise}\%)$, where $lb_{noise}$ and $ub_{noise}$ denote the lower and upper bounds respectively. During training of the CNN, we consider different scenarios for these bounds and determine appropriate levels by cross-validation. In Figure 9, we show noise realizations sampled from noise distribution 3, which was found to be consistent with $\boldsymbol{d}_{obs}$ as shown below.

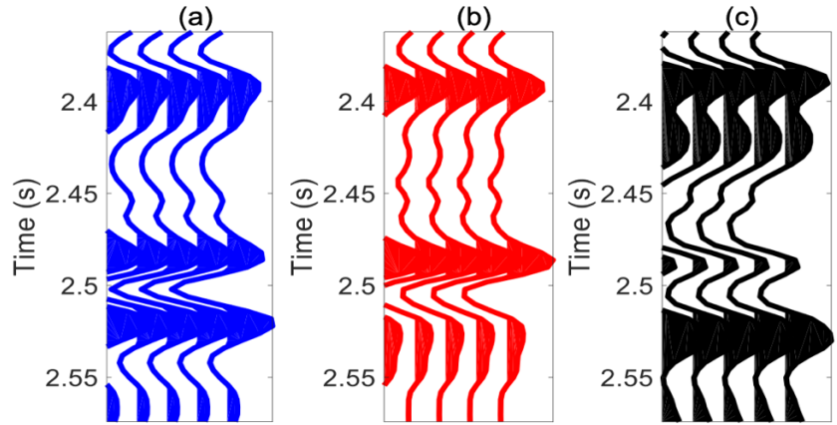

Figure 10: Residual traces calculated from the error between synthetic and real traces for (a) post-stack, (b) near-stack and (c) far-stack seismic data.

## Prior falsification

Specification of the prior model required making several subjective decisions such as the nature and support of the prior probability distributions and type of wavelets. To ensure the validity of these decisions, we employ prior falsification analysis which compares statistics of the data samples drawn from the prior with the observed data. For establishing consistency of the underlying geological conceptual model and other modeling parameters with seismic data, comparison of the global features present in the data is desirable as opposed to a local trace-by-trace comparison. Scheidt et al. (2015) propose extracting features from seismic data by discrete wavelet transform (DWT) to give the approximation and multi-resolution detail coefficients (Mallat, 1989; Mallat, 1999). The wavelet transform representation of 3D cubes constitutes of the approximation coefficients at the coarsest resolution and detail coefficients at multiple higher resolutions. Global comparison between any two data volumes is performed by comparing kernel density estimates (KDEs) of global histograms of each individual DWT coefficients.

The analysis is performed with 500 prior samples of $\boldsymbol{d} = [\boldsymbol{d}^{post}, \boldsymbol{d}^{near}, \boldsymbol{d}^{far}]^T$ and observed data $\boldsymbol{d}_{obs} = [\boldsymbol{d}_{obs}^{post}, \boldsymbol{d}_{obs}^{near}, \boldsymbol{d}_{obs}^{far}]^T$. We decompose each partial stack seismic volume with a 5-level 3D DWT, performed using Daubechies least asymmetric wavelet bases (Daubechies, 1992). To compare how dissimilar the statistics of the two samples $\boldsymbol{d}^i$ and $\boldsymbol{d}^j$ are, a quantitative measure of dissimilarity is required. This is computed as $D_{ij}^{DWT} = $

$$\sqrt{\sum_{s=1}^{3} \sum_{wc=1}^{N} \left(\frac{D_{ij,s}^{wc}}{\sigma(D_{ij,s}^{wc})}\right)^2}.$$ Here, $D_{ij,s}^{wc}$ refers to the dissimilarity for wavelet coefficient type $wc$ and

index $s$ iterates through the three seismic stack volumes. The notation $wc$ refers to specific approximation or detail DWT coefficient type being evaluated. Note that $D_{ij}^{wc}$ needs to be normalized by its standard deviation $\sigma(.)$, calculated empirically from the samples, since the scale of coefficient values across different $wc$ may vary considerably. We assign $D_{ij}^{wc}$ as the Jensen-Shannon divergence $\frac{1}{2} D^{KL}(f_i^{wc} || f_{ij}^{wc}) + \frac{1}{2} D^{KL}(f_j^{wc} || f_{ij}^{wc})$, where $f_i^{wc}$ and $f_j^{wc}$ are the histogram KDEs of $wc$ coefficients of $\boldsymbol{d}_i$ and $\boldsymbol{d}_j$ respectively, $D^{KL}(.)$ is the Kullback-Leibler (KL) divergence and $f_{ij}^{wc} = \frac{1}{2}(f_i^{wc} + f_j^{wc})$. The KL divergence between two probability distributions $f_1(.)$ and $f_2(.)$ is given as $D^{KL}(f_1 || f_2) = \int_x f_1(x) \log\left(\frac{f_1(x)}{f_2(x)}\right) dx$. Outlier detection is subsequently accomplished in two steps. In the first step, the data samples are projected into a lower-dimensional space by multi-dimensional scaling (MDS; Borg and Groenen, 1997). MDS is particularly useful since it performs dimension reduction by preserving any desired pair-wise distance measure between the samples. We assign this distance measure to be as $D_{ij}^{DWT}$. Thus, if two samples are similar in terms of their wavelet coefficient histograms, they will be projected to nearby locations in the MDS space. In the second step, the coordinates of the samples in the MDS space are used to perform the Mahalanobis-distance based outlier detection as described in the methodology section. Dimensionality of the MDS space is chosen to be 2, explaining about 90% of the variability observed in the original uncompressed dimensions.

We perform the falsification analysis with and without corruption of data samples by additive noise $\boldsymbol{\epsilon}$. Figure 11 shows the Mahalanobis-distance based outlier detection results for the latter scenario. In the case where the Walden and White (1998) wavelet is used in $g_{\text{WP}}(.)$, $\boldsymbol{d}_{obs}$ falls above the threshold $\tau$, signaling inconsistency between observed data and the synthetic prior samples. On the other hand, the choice of the statistical wavelet cannot be falsified as the outlier detection algorithm considers the prior samples and $\boldsymbol{d}_{obs}$ to belong to the same statistical population. The above analysis also indicates that the neither the geological prior model nor the rock physics model is falsified, and consequently can be used to generate the training data. This exercise is subsequently repeated to determine which of the three distributions considered for noise $\boldsymbol{\epsilon}$ is applicable (Figure 12). Results are shown for $lb_{noise} = 0.5\%$ and $ub_{noise} = 30\%$. It can be seen that both noise distributions 1 and 2 are falsified. In contrast, modeling horizontal and vertical correlations in the additive noise is consistent with $\boldsymbol{d}_{obs}$. To further validate our analysis, we also compare the global distributions of amplitudes of the synthetic and real data (Figure 13). We see a good overlap between the synthetic training data and real data distributions. Thus, noise distribution 3 is not falsified, and may be used during generation of the training examples. Neither the noise-free synthetic prior nor the prior with noise distribution 3 are falsified and hence can be used to generate the synthetic training set. However as shown later the noise-free case leads to overfitting and poor performance with real data since the forward modeling is not perfect and real data has noise. Noise with the right distribution (not falsified) has to be added to the synthetically generated training set to regularize the training, avoid overfitting, and give better performance with real data.

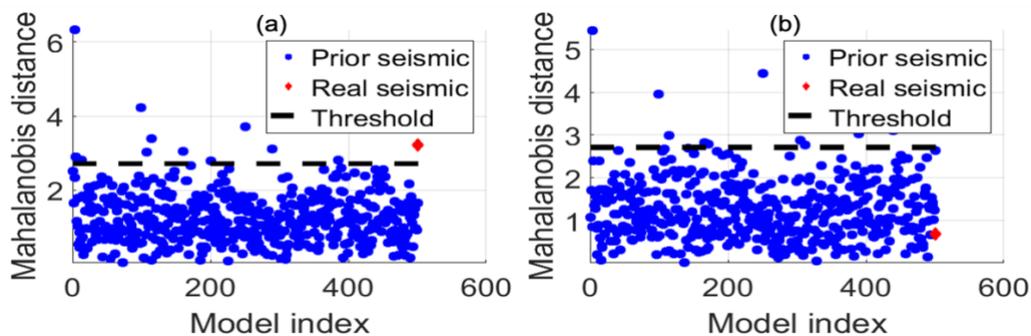

Figure 11: Robust Mahalanobis distances of 500 prior samples of forward modeled seismic data, without additive noise, and real seismic data. Shown are the cases when seismic data is modeled with (a) Walden and White (1998) wavelet and (b) statistical wavelet. Shown in dotted black line is the threshold value for outlier detection.

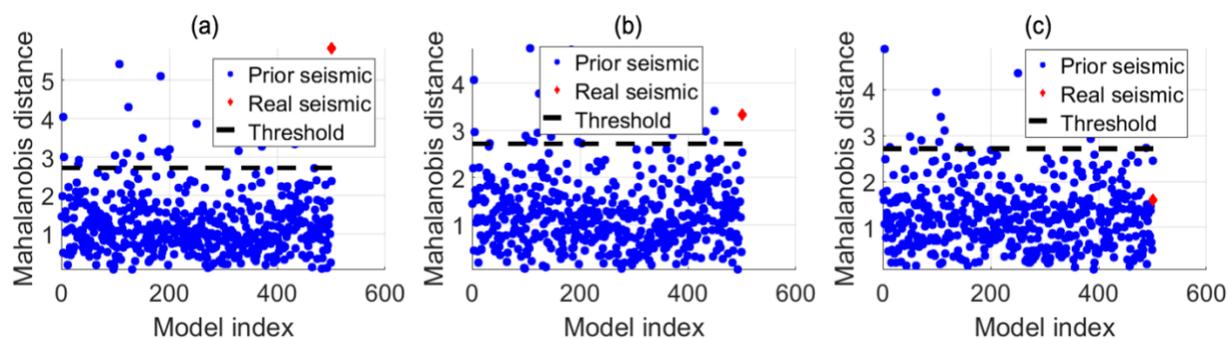

Figure 12: Robust Mahalanobis distances of 500 prior samples of forward modeled seismic data, with additive noise, and real seismic data. Shown are the cases when additive noise is sampled from (a) noise distribution 1, (b) noise distribution 2 and (b) noise distribution 3. Shown in dotted black line is the threshold value for outlier detection.

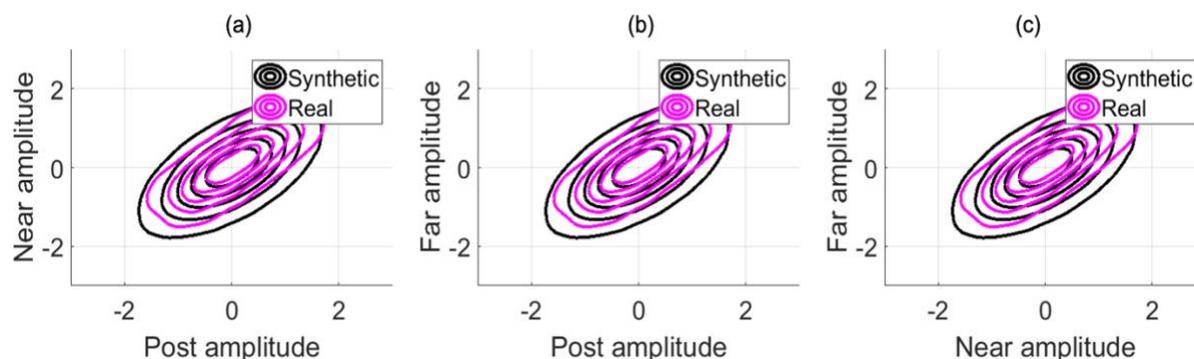

Figure 13: Bivariate kernel density estimates of the global distributions of (a) post stack vs. near stack, (b) post stack vs. far stack and (c) near stack vs. far stack seismic amplitudes. The synthetic data are generated with noise distribution 3.

**Facies estimation with 3D CNN and real seismic data**

In this section, we present the results of facies classification by 3D CNNs using seismic amplitude data as features. Note that we pose the classification problem for prediction of three facies: gas sands, brine sands and non-reservoir facies, as opposed to the well-scale facies prior model containing six different facies. Splays, levees, thin beds and shale facies were lumped together into non-reservoir facies as analysis of direct log measurements showed significant overlap between these facies in the P-impedance and $V_P - V_S$ domain ([Figure 4](#)). We further quantify this observation by performing Bayesian classification (Duda et al., 2000; Avseth et al., 2005) of the log measurements using P-impedance and $V_P - V_S$ as features. The corresponding classification confusion matrix is shown in [Table 4](#). For a classification problem with $k$ classes, the confusion matrix is a $k \times k$ matrix, the $ij^{th}$ entry of which gives the conditional probability that the true class is $i$ given that predicted class by the classifier is $j$. [Table 4](#) shows that splays, levees and thin beds are primarily misclassified as shales, thus indicating that seismic data will not be able to resolve these non-reservoir facies from each other. In [Figure 4](#), we show how the density estimates in the P-impedance and $V_P - V_S$ domain update after lumping all non-reservoir facies together. The confusion matrix for Bayesian classification of the three facies with P-impedance and $V_P - V_S$ features is shown in [Table 5](#).

Table 4: Confusion matrix for Bayesian classification of the facies with P-impedance and $V_P$-$V_S$ ratio as features.

| | | Predicted facies | | | | | |
|---|---|---|---|---|---|---|---|
| | | Brine sand | Splay | Shale | Levee | Thin bed | Gas sand |
| **True facies** | **Brine sand** | 0.77 | 0.1 | 0.1 | 0 | 0.01 | 0.02 |
| | **Splay** | 0.12 | 0.2 | 0.47 | 0.07 | 0.03 | 0.11 |
| | **Shale** | 0 | 0 | 0.96 | 0.03 | 0 | 0.01 |
| | **Levee** | 0.01 | 0 | 0.82 | 0.1 | 0 | 0.07 |
| | **Thin bed** | 0 | 0 | 0.98 | 0.02 | 0 | 0 |
| | **Gas sand** | 0.02 | 0 | 0.09 | 0.08 | 0 | 0.81 |

Table 5: Confusion matrix for Bayesian classification of brine sand, gas sand and non-reservoir facies with P-impedance and $V_P$-$V_S$ ratio as features.

| | | Predicted facies | | |
|---|---|---|---|---|
| | | Brine sand | Non-reservoir | Gas sand |
| **True facies** | **Brine sand** | 0.74 | 0.24 | 0.02 |
| | **Non-reservoir** | 0.03 | 0.93 | 0.04 |
| | **Gas sand** | 0.01 | 0.25 | 0.74 |

*CNN architecture, training and overfitting*

Since $\boldsymbol{h}$ is a 3D cube while $\boldsymbol{d}$ is a 4D numerical array, we are seeking to solve a spatially dense classification problem. CNNs model the relationship between inputs and outputs through multiple hidden convolutional layers, each layer having multiple filters. Both convolutional layers and filters possess spatial dimensions such as width, height and depth. The size of the filters is kept smaller than the input layer and thus they operate on local regions of the input.

Each filter may activate according to distinctive spatial features present in the input. For any hidden layer, the output at a single pixel of the subsequent layer is given as:

$$y^{c,R} = g\left(\sum_i w_i^c x_i^R + b^c\right). \tag{7}$$

Here, $w_i^c$ denotes a learnable weight coefficient of the convolutional filter indexed by $c$, with $c = \{1,..,n_c\}$ and $i = \{1,..,n_k\}$. The total number of filters for the layer is denoted by $n_c$, while $n_k$ denotes the number of filter coefficients in each filter. The bias term is denoted by $b^c$ and $g(.)$ denotes a non-linear activation function. Frequently used activation functions include the sigmoid function, $g(x) = 1/(1 + e^{-x})$, the hyperbolic tangent function, $g(x) = (e^x - e^{-x})/(e^x + e^{-x})$ and the rectified linear unit (ReLU), $g(x) = \max(0, x)$ We use $x_i^R$ to denote the $i^{th}$ pixel within the region indexed with $R$. To compute output at an adjacent pixel, the filter is spatially translated by a certain stride to operate on an adjacent region of the input. The maximum value that $R$ can take will vary depending on the stride parameters set during the filter translation operations. CNNs scale-up efficiently to high-dimensional data due to their usage of spatially invariant local filters. In addition to convolutional layers, pooling layers, such as max and average pooling, and upsampling layers, such as transposed convolutional layers, are commonly employed in the CNN architectures. Difference in the input and output dimensions may be reconciled using these layers. Dumoulin and Visin (2016) provide a detailed treatment of the arithmetic of these layers.

By stacking multiple convolutional layers, each featuring multiple kernels, CNNs possess the ability to extract multiple complex feature representations from the input. Even though every individual network layer constitutes of activations from local patches of the preceding layer,

CNNs can aggregate features across multiple spatial scales by progressively increasing the receptive field in the deeper layers of the network. Here, the receptive field of any hidden layer refers to the effective spatial size of the first input layer that the hidden layer is informed by (Yu and Koltun, 2016). Several CNN architectural designs have been proposed to achieve this in practice. We consider strategies from dense image classification problems, also known as semantic image segmentation (Long et al., 2015; Yu and Koltun, 2016; Chen et al., 2017), because of their similarity to the objective of this paper. A popular CNN architecture for semantic segmentation is the dilated-convolutional architecture (Yu and Koltun, 2016). The core ingredient is the usage of dilatational convolutional layers for exponentially increasing the receptive field, dispensing with the necessity of downsampling and upsampling the hidden layers as is common in encoder-decoder architectural designs (Long et al., 2015). In dilatational layers, the convolutional filter is padded with zeros in between the non-zero filter coefficients. This allows boosting the receptive field without affecting the spatial size of the network layers or the number of learnable parameters.

The dilated CNN network architecture we employ is shown in Table 6. It was designed to take as input the post-stack, near-stack and far-stack seismic cubes, stacked as a 4D array of dimensions $60 \times 100 \times 100 \times 3$. Note that we do not apply any additional normalization to the input data apart from the global normalization described in the seismic forward modeling section. A key difference with conventional semantic segmentation architectures is that the input and output domains are not the same and have different sizes along the time or depth dimension. The time dimension of seismic grid is upsampled to the depth dimension of the reservoir grid using transposed convolutional layers. Note especially the exponential increase of the dilatation factors in the deeper convolutional layers. Filters are dilated such that the receptive field of the output layer is at least equivalent to the dimensionality of the input layer. For instance, using the dilatation scheme shown in [Table 6], receptive fields of the output volume along $x$ and $y$ axes equal 131. This ensures that all large-scale horizontal features present in the input seismic grid fall within the field of view of each pixel in the final layer. We use batch-normalization (Ioffe and Szegedy, 2015) layers after all regular and transposed convolutional layers. ReLU is used as non-linear activation function in all the layers except the last, where the softmax function is used. This models the marginal conditional distribution $f_{\mathrm{CNN}}(h_i = j|\boldsymbol{d})$ of any pixel $i$ of the reservoir grid as the single-trial multinomial distribution, with number of possible outcomes equal to the count of facies categories $k$. The multinomial distribution is parameterized by the probability of each class

$$f_{\mathrm{CNN}}(h_i = j|\boldsymbol{d}) = \frac{e^{\eta_i^j}}{\sum_{l=1}^{k} e^{\eta_i^l}}, \forall \{j = 1, \ldots k\}, \tag{8}$$

modeled using the softmax activation function. Here, $\eta_i^j$ denotes the output value of final layer at pixel $i$ and channel $j$. The CNN output layer thus has $k = 3$ channels for the three facies under consideration. We assume conditional independence of all the reservoir pixels conditioned to seismic data and express the approximation to the joint conditional distribution as

$$f_{\mathrm{CNN}}(\boldsymbol{h}|\boldsymbol{d}) = \prod_{i=1}^{n^h} \hat{f}(h_i|\boldsymbol{d}). \tag{9}$$

The above assumption is particularly useful as the gradient of each pixel in the objective function with respect to the network parameters can be computed independently. This allows training the CNN with conventional gradient-based optimization methods used for training of deep learning networks.

Table 6: The CNN architecture used in the real case study. 'Conv3D': 3D convolutional layer, 'TranspConv3D': 3D transposed convolutional layers and 'ReLU': rectified linear unit. Batch-normalization used after all 'Conv3D' and 'TranspConv3D' layers.

| Layer type | Filter size | # of filters | Conv type | Strides | Dilatation factor | Activation | Output shape |
|---|---|---|---|---|---|---|---|
| Input | | | | | | | $60 \times 100 \times 100 \times 3$ |
| Conv3D | $3 \times 3 \times 3$ | 32 | Same | $1 \times 1 \times 1$ | $1 \times 1 \times 1$ | ReLU | $60 \times 100 \times 100 \times 32$ |
| Conv3D | $3 \times 3 \times 3$ | 32 | Same | $1 \times 1 \times 1$ | $1 \times 1 \times 1$ | ReLU | $60 \times 100 \times 100 \times 32$ |
| Transp Conv3D | $33 \times 1 \times 1$ | 32 | Valid | $1 \times 1 \times 1$ | $1 \times 1 \times 1$ | ReLU | $92 \times 100 \times 100 \times 32$ |
| Conv3D | $3 \times 3 \times 3$ | 32 | Same | $1 \times 1 \times 1$ | $2 \times 2 \times 2$ | ReLU | $92 \times 100 \times 100 \times 32$ |
| Conv3D | $3 \times 3 \times 3$ | 32 | Same | $1 \times 1 \times 1$ | $4 \times 4 \times 4$ | ReLU | $92 \times 100 \times 100 \times 32$ |
| Transp Conv3D | $68 \times 1 \times 1$ | 32 | Valid | $2 \times 1 \times 1$ | $1 \times 1 \times 1$ | ReLU | $250 \times 100 \times 100 \times 32$ |
| Conv3D | $3 \times 3 \times 3$ | 32 | Same | $1 \times 1 \times 1$ | $8 \times 8 \times 8$ | ReLU | $250 \times 100 \times 100 \times 32$ |
| Conv3D | $3 \times 3 \times 3$ | 32 | Same | $1 \times 1 \times 1$ | $16 \times 16 \times 16$ | ReLU | $250 \times 100 \times 100 \times 32$ |
| Conv3D | $3 \times 3 \times 3$ | 32 | Same | $1 \times 1 \times 1$ | $32 \times 32 \times 32$ | ReLU | $250 \times 100 \times 100 \times 32$ |
| Conv3D | $3 \times 3 \times 3$ | 3 | Same | $1 \times 1 \times 1$ | $1 \times 1 \times 1$ | Softmax | $250 \times 100 \times 100 \times 3$ |

The CNN (Table 6) was implemented in Tensorflow deep learning framework. We employ following two metrics to gauge the network's prediction performance on the different evaluation sets such training and validation sets.

1. Voxel-wise classification accuracy: Defined as $\frac{\sum_{l=1}^{n} \sum_{i=1}^{n_h} \mathbb{I}\{\hat{h}_i^{(l)} = h_i^{(l)}\}}{\sum_{l=1}^{n} \sum_{i=1}^{n_h} 1}$, this metric calculates accuracy averaged across all $n_h$ pixels in the output grid and all $n$ examples in a particular evaluation set. Here, $\mathbb{I}(.)$ denotes the indicator function. The facies category at any voxel $\hat{h}_i$ is calculated by applying a hard threshold of 0.5 to the probabilities $\hat{f}(h_i | \boldsymbol{d}_i)$ predicted by the CNN.

2. AUC-ROC curves: For classifiers giving continuous valued outputs, such as the CNN classifier, applying a single hard threshold to estimate the class labels might not be effective in evaluating the classifier performance. Fawcett (2006) introduced receiver operating characteristic (ROC) curves to address this limitation. ROC graphs are 2D graphs plotting the true positive rate against the false positive rate for every facies category under consideration. For a facies category $k$, the terms positives and negatives are used to indicate the presence or absence of the category at a voxel. If the true class at any given voxel is $k$ and is correctly classified by the classifier, it is counted as a true positive instance. If the classifier incorrectly classifies a voxel as class $k$, it is counted as a false positive instance. The true positive rate for class $k$ is given as $\frac{True\ positives}{Total\ positives}$, while the false positive rate given as $\frac{False\ positives}{Total\ negatives}$. The ROC curve of true and false positive rates is computed by varying the thresholds applied to CNN class probabilities to estimate facies classes. The area under the curve (AUC) of the ROC curve may then be used to evaluate the network performance. A perfect classifier will always have a true positive rate of 1, leading to AUC-ROC metric value of 1 for all classes. For a random classifier, the true and false positive rates will always be equal, resulting in AUC-ROC metric value of 0.5.

To demonstrate the effect of overfitting to the synthetic training data distribution, we first formulate evaluation sets by sampling from $f_{\text{synth}}(\boldsymbol{h}, \boldsymbol{m}, \boldsymbol{d})$ without any additive noise $\boldsymbol{\epsilon}$. To be clear, we denote this distribution as $f_{\text{synth}}^{\text{noise-free}}(\boldsymbol{h}, \boldsymbol{m}, \boldsymbol{d})$. 2400 samples were generated from this distribution and subsequently split into training, validation and test sets of sizes 2000, 200 and 200 respectively. The network was trained using the cross-entropy loss function and voxel-wise classification accuracy metric for 40 epochs (~ 24 hours run time) with Adam optimizer (Kingma and Ba, 2015) on a machine with 128 GB RAM and two 32GB Tesla V100 GPUs. We used $\ell2$ regularization (Bishop, 2006) to prevent any potential overfitting to the training set. After 40 epochs of training, the CNN learns to make predictions with high classification accuracies (~90%) for the training and validation sets. Similar classification accuracies were obtained on the test set. In Figure 14 and Figure 15, we compare the true facies against CNN predictions along depth and cross sections of a test set model. It can be observed that the CNN has almost perfectly learned to predict the complex geometries and curvilinear geological features of the channel objects. The spatial locations of the channels and depths to the GWC have also been predicted with high accuracy. Figure 16 shows the corresponding ROC curves for CNN classification on this example. We use the sum of the AUC-ROC curves for all three facies as a metric for evaluating classification quality. The AUC-ROC metric has a value of 2.84 for this test set example, indicating a high degree of agreement of the true model with the predicted class probabilities. Recall that the perfect classifier will have AUC-ROC value of 1 for each facies, leading to cumulative metric value of 3 for all facies.

Even though the network seems to have generalized to random samples from $f_{\text{synth}}^{\text{noise-free}}(\boldsymbol{h}, \boldsymbol{m}, \boldsymbol{d})$, we found that the network performs poorly with real data $\boldsymbol{d}_{obs}$. Specifically, $\boldsymbol{d}_{obs}$ was used as input to the CNN and predicted class probabilities along the wellbores of all six wells in the reservoir grid were extracted. Predicted probabilities were compared with the legacy petrophysical interpretations of the facies at the wells. Note that these interpretations were available to us at the well-log scale. In order to account for the discrepancy between the well-log and seismic resolution, we performed upscaling of the facies logs to a scale of approximately 1/10 of the seismic wavelength (Avseth et al., 2005). Corresponding ROC curves are shown in Figure 16. We see that the network performs poorly with a total AUC-ROC metric value of 1.9. Earlier we established that (1) $f_{\text{synth}}^{\text{noise-free}}(\boldsymbol{h}, \boldsymbol{m}, \boldsymbol{d})$ is statistically consistent with $\boldsymbol{d}_{obs}$ during the prior falsification analysis, and (2) the CNN is not overfitting to the training set. Hence, a possible reason could be that the network is overfitting to the distribution $f_{\text{synth}}^{\text{noise-free}}(\boldsymbol{h}, \boldsymbol{m}, \boldsymbol{d})$. This would cause the network to make erroneous predictions with $\boldsymbol{d}_{obs}$ since modeled $f_{\text{synth}}^{\text{noise-free}}(\boldsymbol{h}, \boldsymbol{m}, \boldsymbol{d})$ contains modeling imperfections and real data is noisy.

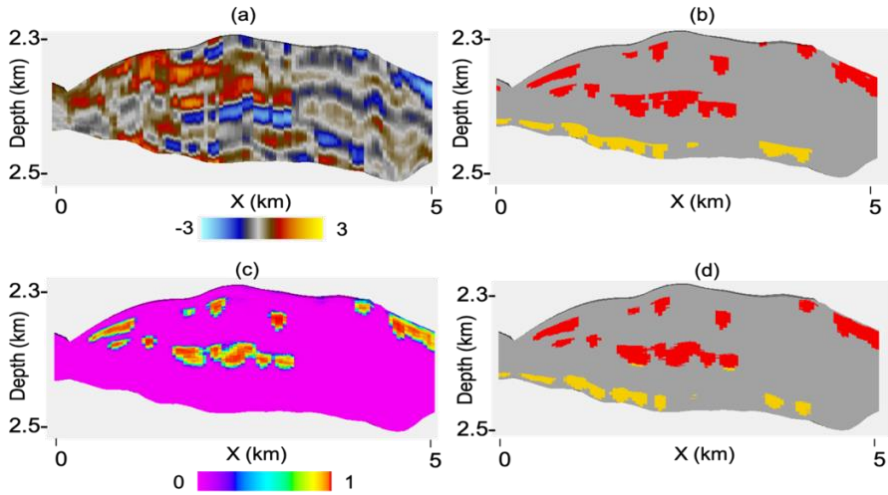

Figure 14: Cross-sections of (a) synthetic post-stack seismic data, (b) true facies section, (c) predicted gas sand probability and (d) most likely facies prediction. In the facies sections, red, yellow and gray represent gas sands, brine sands and non-reservoir facies respectively.

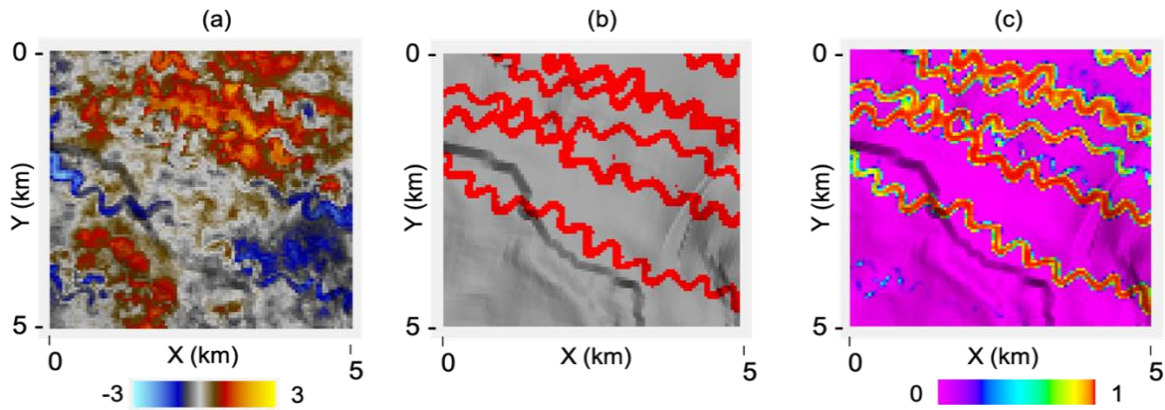

Figure 15: Horizon sections of (a) synthetic post-stack seismic data, (b) true facies section and (c) predicted gas sand probability. In the facies section, red, yellow and gray represent gas sands, brine sands and non-reservoir facies respectively.

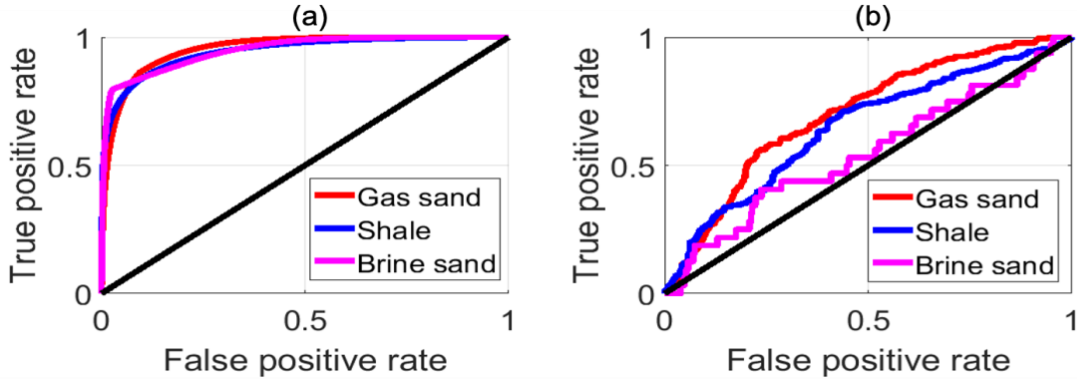

Figure 16: ROC curves for the CNN classifier trained with synthetic examples without any additive noise. ROC curves are shown for facies classification (a) on a synthetic test set example with synthetic seismic data as input features and (b) at wells with real seismic data as input features. Shown in black is the ROC curve for the random classifier.

*Making reliable predictions with real data*

As motivated in the methodology section, we employ cross-validation with real data and additive noise modeling to make reliable predictions with real data. We employed the validation set $\mathfrak{D}_{\text{real}} = \{(\mathbb{S}[\boldsymbol{h}_{\text{true}}], \boldsymbol{d}_{obs})\}$ for cross-validation during training. We performed cross-validation with five out of the six wells in the study area. Well 5 is kept completely blind. We evaluated both the evaluation metrics introduced previously and found the AUC-ROC metric to be effective in producing reliable results. To summarize the cross-validation process, we obtain facies predictions from the CNN with $\boldsymbol{d}_{obs}$ after each training epoch and calculate the AUC-ROC metric on $\mathfrak{D}_{\text{real}}$. We retain the network weights at the epoch with the highest AUC-ROC metric value on the well validation set. For the training experiments described below, it was found that the optimal AUC-ROC metric on $\mathfrak{D}_{\text{real}}$ was generally obtained within the first three training epochs.

We incorporated the additive noise distribution $\boldsymbol{\epsilon}$ into training data distribution $f_{synth}(\boldsymbol{h}, \boldsymbol{m}, \boldsymbol{d})$. Noise distribution 3, found to be statistically consistent with $\boldsymbol{d}_{obs}$ during prior falsification, was sampled to corrupt the training examples. To determine the right amount of noise that is effective against overfitting, we created several sets of the synthetic training data, modeled with various levels of signal-noise ratio. Specifically, the additive noise upper and lower bounds, $lb_{noise}$ and $ub_{noise}$, were systematically varied as shown in Table 7. For each signal-to-noise ratio scenario considered, we trained the CNN on the corresponding training set and performed cross-validation with $\mathfrak{D}_{\mathrm{real}}$. In Table 7, we show the ROC-AUC metric calculated on $\mathfrak{D}_{\mathrm{real}}$. It can be observed that prediction quality at the wells deteriorates with too small and too large levels of noise. Low level of noise is not sufficient in preventing overfitting while too much noise restricts the ability to learn a useful model. The best ROC-AUC metric is obtained for $lb_{noise} = 30\%$ and $ub_{noise} = 70\%$ and subsequent results are shown for this scenario.

In Figure 17 and Figure 18, we plot the CNN predictions along different sections with the corresponding section from the post-stack seismic data sampled into the reservoir grid. It can be seen that the CNN seems to preserve the spatial continuity of curvilinear channel-like features that are evident in the data. The variance map, computed as the variance of the multinomial distribution modeled by the last layer of the CNN, shows the corresponding uncertainty in the predictions. Most of the uncertainty is localized around the predicted channel-like features. In Figure 19, we compare the predictions of the CNN against petrophysical facies interpretations at wells 2, 5 and 6. We see that for well 2, the network has predicted with high probability the occurrences of the two gas sand beds in the well. Also, note the high variance in the predictions, especially around the edges of the two beds. Note that well 6 was kept blind during the prior building process but used for cross-validation, while well 5 was kept completely blind. For both these wells, the network identifies the presence of the sand beds. There exist small depth mismatches in the sand bed locations in the wells and predictions. This was primarily because seismic-well tie was performed only using well 1 due to non-availability of the complete set of $V_P, V_S$ and $\rho_b$ logs at remaining wells. We visualize the CNN classification performance in Figure 20 with ROC curves obtained using CNN classifications at all wells. The curves have an AUC-ROC metric value of 2.58. We also show the ROC curves obtained from Bayesian facies classification at wells using log-scale features of P-impedance and $V_P - V_S$. The corresponding AUC-ROC metric has a value of 2.80. The drop in classification performance with CNNs is expected due to the (1) limited resolution of seismic data, (2) complex data noise signatures and unmodeled geological and geophysical processes that cannot be captured with Gaussian additive noise distributions and (3) absence of seismic-well ties at some wells due to missing logs.

Table 7: AUC-ROC metrics obtained with different additive noise distributions.

| $lb_{noise}\%$ | $ub_{noise}\%$ | AUC-ROC metric evaluated on $\mathfrak{D}_{\mathbf{real}}$ |
|---|---|---|
| 0 | 0 | 2.01 |
| 0.5 | 30 | 2.18 |
| 0.5 | 50 | 2.52 |
| 0.5 | 70 | 2.50 |
| 0.5 | 90 | 2.56 |
| 30 | 70 | 2.59 |
| 50 | 90 | 2.07 |
| 100 | 100 | 2.19 |

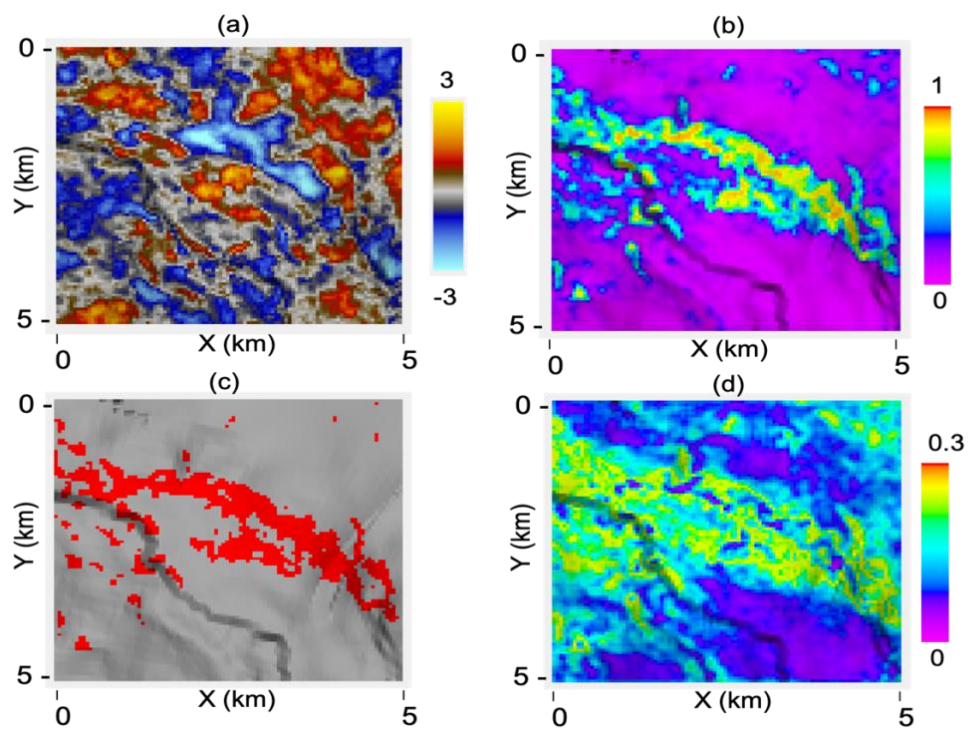

Figure 17: (a) Horizon slices of (a) field post-stack seismic data, (b) predicted gas sand probability, (c) most-likely facies model and (d) variance of gas sand predictions. In the facies section, red, yellow and gray represent gas sands, brine sands and non-reservoir facies respectively.

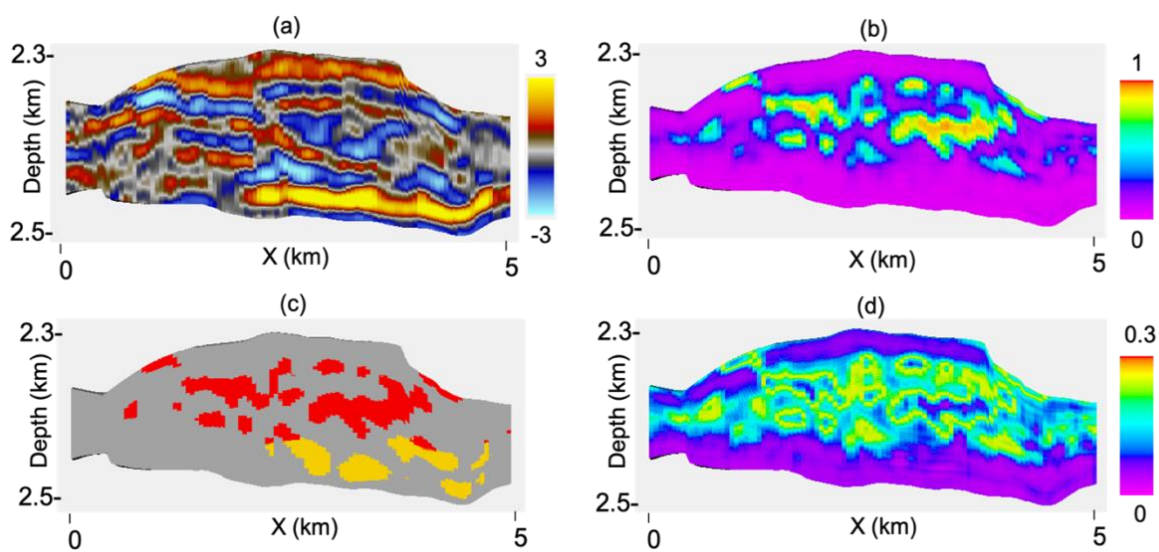

Figure 18: Cross sections of (a) field post-stack seismic data, (b) predicted gas sand probability, (c) most-likely facies model and (d) variance of gas sand predictions. In the facies section, red, yellow and gray represent gas sands, brine sands and non-reservoir facies respectively.

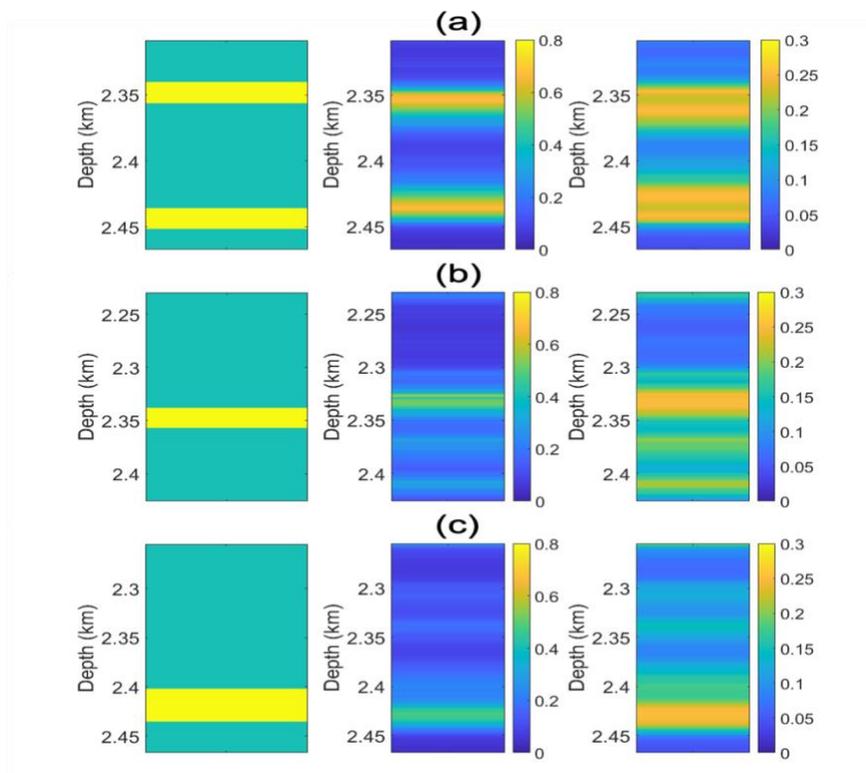

Figure 19: Legacy petrophysical facies interpretations (left; sand: yellow, shale: green), CNN predicted gas sand probability (middle) and prediction variance (right) for (a) well 2, (b) 5 and (c) 6.

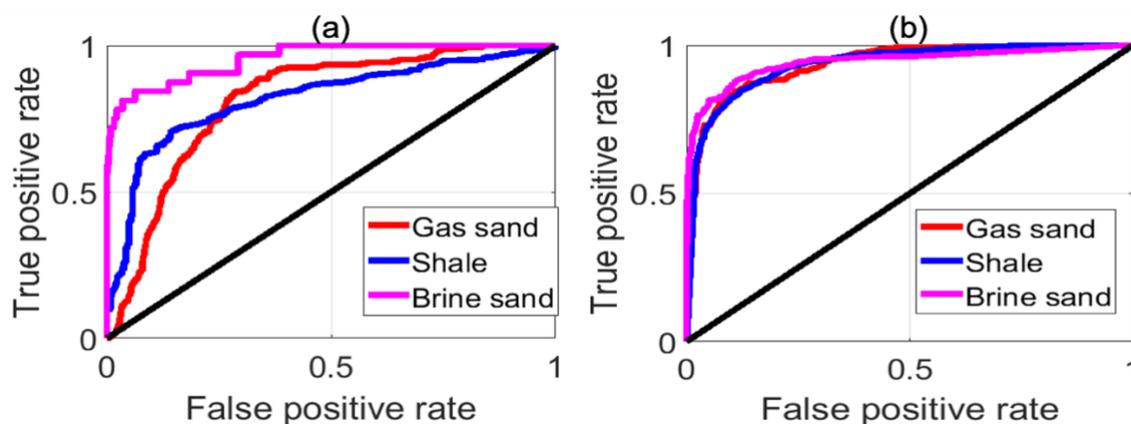

Figure 20: ROC curves for facies classification at all the wells in the study area obtained from the (a) CNN classifier with real seismic data as input features and (b) Bayesian classifier with well log-based input features. Shown in black is the ROC curve for the random classifier.

DISCUSSION

In this section, we discuss limitations and advantages associated with proposed approach. Creation of synthetic training data requires specification, modeling, and validation of the prior geologic uncertainty. Such an exercise could potentially be time intensive, as it might require iterative modifications to the prior model until the consistency criteria with real data is met. This might especially be an issue in frontier exploration settings, where limited information about the subsurface geology is available. However, once a consistent prior uncertainty model is established, creating the training data is straightforward to accomplish since Monte-Carlo sampling is employed to sample from the statistical distribution. This aspect makes the methodology significantly more efficient than iterative data-matching approaches, such as Markov chain Monte-Carlo based Bayesian inversion, in high-dimensional settings. Also, note that proposed method is not coupled to any specific geostatistical prior model and thus any desired modeling algorithm can be incorporated. We showed that CNNs trained on perfectly modeled noise-free data were prone to overfitting to the prior model, with poor results on the real data. A main finding of this paper was that additive noise modeling aided in boosting the prediction quality at wells. However, parametric noise distributions, such as Gaussian distributions, will not be able to capture all complex signatures of data noise and modeling imperfections. Consequently, the statistics of the real data might contain small deviations from the training data distribution in some cases. We proposed tuning the noise level to an appropriate value by cross-validation. Further research is needed on strategies to ensure that the CNNs becomes less sensitive to such discrepancies in the input distribution.

# CONCLUSIONS

Creating training datasets for DL in subsurface problems is challenging due to absence of repeated measurements of data and target variables. In this paper, we presented a real case application of reservoir facies classification in an offshore deltaic reservoir with deep 3D CNNs and 3D partial angle-stack seismic data, and proposed strategies to create and validate synthetic labeled training data. Training data were created by specifying and sampling probability distributions modeling the prior geological uncertainty on reservoir properties and forward modeling the seismic data. While deep 3D CNNs demonstrated excellent ability to approximate the underlying statistical distributions with high fidelity, we found that their prediction performance deteriorates with real data if the synthetic training data is not consistent with the real data. We found the approach of prior falsification effective in modifying the prior probability distributions to establish statistical consistency between the synthetic and real data. We found that CNNs are prone to overfitting to synthetic training data distribution, which may lead to erroneous predictions with real data even if the network is demonstrating high accuracy on blind synthetic test data. Early stopping of the training with real data and corrupting the training data with noise random variables, capturing modeling imperfections and data noise, were found to be critical in making the CNN trained on synthetic data efficacious with real data.

REFERENCES

Aleardi, M., and F. Ciabarri, 2017, Assessment of different approaches to rock-physics modeling: a case study from offshore Nile Delta: Geophysics, **82**, no. 1, MR15–MR25, https://doi.org/10.1190/geo2016-0194.1

Aleardi, M., F. Ciabarri, and R. Calabrò, 2018, Two-stage and single-stage seismic-petrophysical inversions applied in the Nile Delta: The Leading Edge, **37**, 510–518, https://doi.org/10.1190/tle37070510.1

Aki, K., and P. G. Richards, 1980, Quantitative seismology: W. H. Freeman & Co.

Avseth, P., T. Mukerji, and G. Mavko, 2005, Quantitative Seismic Interpretation: Cambridge University Press.

Bishop, C. M., 1995, Training with Noise is Equivalent to Tikhonov Regularization: Neural Computation, **7**, no. 1, 108–116, https://doi.org/10.1162/neco.1995.7.1.108

Bishop, C. M., 2006, Pattern recognition and machine learning: Springer-Verlag.

Borg, I., and P. Groenen, 1997, Modern Multidimensional Scaling: Theory and Applications: Springer.

Caers, J., 2005, Petroleum Geostatistics: Society of Petroleum Engineers.

Chen, L. C., G. Papandreou, I. Kokkinos, K. Murphy, and A. L. Yuille, 2017, DeepLab: Semantic image segmentation with deep convolutional nets, atrous convolution, and fully connected CRFs: IEEE transactions on pattern analysis and machine intelligence, **40**, no. 4, 834-848.

Cross, N. E., A. Cunningham, R. J. Cook, A. Taha, E. Esmaie, N. E. Swidan, 2009, Three-dimensional seismic geomorphology of a deep-water slope-channel system: The Sequoia field, offshore west Nile Delta, Egypt: AAPG Bulletin, **93,** no. 8, 1063–1086, https://doi.org/10.1306/05040908101

Daubechies, I., 1992, Ten lectures on wavelets: CBMS-NSF conference series in applied mathematics, SIAM Ed.

Das, V., A. Pollack, U. Wollner, and T. Mukerji, 2019, Convolutional neural network for seismic impedance inversion: Geophysics, **84**, 1–66.

Das, V., and T. Mukerji, 2020, Petrophysical properties prediction from prestack seismic data using convolutional neural networks: Geophysics, **85**, N41-N55.

Deutsch, C. V., and A. G. Journel, 1998, GSLIB: Geostatistical Software Library and User's Guide: Oxford University Press.

Duda, R. O., P. E., Hart, and D. G. Stork, 2000, Pattern classification: John Wiley and Sons.

Dumoulin, V., and F. Visin, 2016, A guide to convolution arithmetic for deep learnin: arXiv preprint, arXiv:1603.07285.

Fawcett, T., 2006, An introduction to ROC analysis: Pattern Recognition Letters, **27**, 861–874.

Goodfellow, I., Y. Bengio, and A. Courville, 2016, Deep Learning: MIT Press.

Goovaerts, P., 1997, Geostatistics for natural resources evaluation: Oxford University Press.

Ioffe, S., and C. Szegedy, 2015, Batch normalization: accelerating deep network training by reducing internal covariate shift: Proceedings of the 32nd International Conference on Machine Learning, **37**, 448–456.

Kingma, D., and J. Ba, 2015, Adam: a method for stochastic optimization: Proceedings of the 3rd International Conference on Learning Representations.

Krizhevsky, A., I. Sutskever, and G. Hinton, 2012, ImageNet classification with deep convolutional neural networks: Advances in neural information processing systems, **25**, 1097–1105.

Long, J., E. Shelhamer, and T. Darrell, 2015, Fully convolutional networks for semantic segmentation: CVPR.

Mallat, S., 1989, A theory for multiresolution signal decomposition: the wavelet representation: IEEE Transactions on Pattern Analysis and Machine Intelligence, **11**, no. 7, 674-693, doi: 10.1109/34.192463.

Mallat, S., 1999, A Wavelet Tour of Signal Processing: Academic Press.

Mosser, L., O. Dubrule, and M. J. Blunt, 2020, Stochastic Seismic Waveform Inversion Using Generative Adversarial Networks as a Geological Prior: Mathematical Geosciences, **52**, 53–79, https://doi.org/10.1007/s11004-019-09832-6

Ng, A., and M. Jordan, 2002, On discriminative vs. generative classifiers: A comparison of logistic regression and naïve Bayes: Proceedings of the 14th International Conference on Neural Information Processing Systems: Natural and Synthetic, 841-848.

Pradhan, A., and T. Mukerji, 2020a, Seismic inversion for reservoir facies under geologically realistic prior uncertainty with 3D convolutional neural networks: SEG Technical Program Expanded Abstracts, 1516-1520.

Pradhan, A., and T. Mukerji, 2020b, Seismic Bayesian evidential learning: estimation and uncertainty quantification of sub-resolution reservoir properties: Computational Geosciences, **24**, 1121-1140, https://doi.org/10.1007/s10596-019-09929-1.

Pyrcz, M. J., and C. V. Deutsch, 2014, Geostatistical Reservoir Modeling: Oxford University Press.

Rousseeuw, P.J., and K. Van Driessen, 1999, A fast algorithm for the minimum covariance determinant estimator: Technometrics, **41**, 212–223.

Rousseeuw P.J., and B.C Van Zomeren, 1990, Unmasking multivariate outliers and leverage points: Journal of the American Statistical Association, **85**, no. 41, 633-651.

Scheidt, C., C. Jeong, T. Mukerji, and J. Caers, 2015, Probabilistic falsification of prior geologic uncertainty with seismic amplitude data: Application to a turbidite reservoir case: Geophysics, **80**, M89-M12.

Scheidt, C., L. Li, and J. Caers, 2017, Quantifying Uncertainty in Subsurface Systems: Wiley-Blackwell.

Tarantola, A., 2005, Inverse Problem Theory and Methods for Model Parameter Estimation: SIAM.

Tikhonov, A. N., and V. Y. Arsenin, 1977, Solutions of Ill-Posed Problems: V. H. Winston.

Walden, A.T. and R. E. White, 1998, Seismic wavelet estimation: a frequency domain solution to a geophysical noisy input-output problem: IEEE Transactions on Geoscience and Remote Sensing, **36**, 287-297.

Wu, X., L. Liang, Y. Shi, and S. Fomel, 2019, FaultSeg3D: Using synthetic data sets to train an end-to-end convolutional neural network for 3D seismic fault segmentation: Geophysics, **84**, IM35-IM45.

Wu, X., Z. Geng, Y. Shi, N. Pham, S. Fomel, and G. Caumon, 2020, Building realistic structure models to train convolutional neural networks for seismic structural interpretation, Geophysics, **85**, WA27-WA39.

Yi, B.Y., G. H. Lee, H. Kim, H. Jou, D. G. Yoo, B. J. Ryu, and K. Lee, 2013, Comparison of wavelet estimation methods: Geosciences Journal, **17**, 55–63. https://doi.org/10.1007/s12303-013-0008-0

Yang, F., and J. Ma, 2019, Deep-learning inversion: A next-generation seismic velocity model building method: Geophysics, **84**, R583-R599.

Yu F., and V. Koltun, 2016, Multi-scale context aggregation by dilated convolutions: Proceedings of the 4th International Conference on Learning Representations.